\documentclass[]{emulateapj}
\usepackage{psfig,xspace}
\usepackage{natbib}

\shorttitle{Nature of X-ray Surface Brightness Fluctuations in M87}
\shortauthors{P. Ar\'evalo, et al.}

\def\plotthree#1#2#3{\centering \leavevmode
\epsfxsize=.6\columnwidth \epsfbox{#1} \hfil
\epsfxsize=.6\columnwidth \epsfbox{#2} \hfil
\epsfxsize=.6\columnwidth \epsfbox{#3}}
\def\plotfour#1#2#3#4{\centering \leavevmode
\epsfxsize=.99\columnwidth \epsfbox{#1} \hfil
\epsfxsize=.99\columnwidth \epsfbox{#2} \hfil
\epsfxsize=.99\columnwidth \epsfbox{#3}\hfil
\epsfxsize=.99\columnwidth \epsfbox{#4}}

\def\mnras{MNRAS}
\def\apj{ApJ}
\def\aap{A\&A}
\def\apjl{ApJL}
\def\nat{Nature}

\def\chandra{{\it Chandra }}

\begin{document}

\title{On the Nature of X-ray Surface Brightness Fluctuations in M87 }
\author{P. Ar\'evalo\altaffilmark{1}, E. Churazov\altaffilmark{2,3}, I. Zhuravleva\altaffilmark{4,5}, W. R. Forman\altaffilmark{6}, C. Jones\altaffilmark{6}}
\altaffiltext{1}{ Instituto de F\'isica y Astronom\'ia, Facultad de Ciencias, Universidad de Valpara\'iso, Gran Bretana N 1111, Playa Ancha, Valpara\'iso, Chile}
\altaffiltext{2}{ Max-Planck-Institut f\"ur Astrophysik, Karl-Schwarzschild-Strasse 1, 85741 Garching, Germany}
\altaffiltext{3}{ Space Research Institute (IKI), Russian Academy of Sciences, Profsoyuznaya 84/32, 117997 Moscow, Russia}
\altaffiltext{4}{ Kavli Institute for Particle Astrophysics and Cosmology, Stanford University, 452 Lomita Mall, Stanford, California 94305-4085, USA}
\altaffiltext{5}{ Department of Physics, Stanford University, 382 Via Pueblo Mall, Stanford, California 94305-4060, USA}
\altaffiltext{6}{ Harvard-Smithsonian Center for Astrophysics, 60 Garden St., Cambridge, MA 02138}

\label{firstpage}
\begin{abstract}
  X-ray images of galaxy clusters and gas-rich elliptical galaxies show
 a wealth of small-scale features which reflect fluctuations in
 density and/or temperature of the intra-cluster medium. In this paper we study 
 these fluctuations in M87/Virgo, to establish whether sound
 waves/shocks, bubbles or uplifted cold gas dominate the structure. We
 exploit the strong dependence of the emissivity on density and
 temperature in different energy bands to distinguish between these 
 processes. Using simulations we demonstrate that our analysis recovers the leading type of fluctuation even in the presence of projection effects and temperature gradients. 
We confirm the isobaric nature of cool filaments of gas
 entrained by buoyantly rising bubbles, extending to
 7\arcmin\ to the east and south-west, and the adiabatic nature of the
 weak shocks at 40\arcsec\ and 3\arcmin\ from the centre.  For features of $\sim5-10$ kpc, we show that
 the central $4\arcmin \times4\arcmin $ region is dominated by cool
 structures in pressure equilibrium with the ambient hotter gas while
 up to 30\% of the variance in this region can be ascribed to
 adiabatic fluctuations.  The remaining part of the central
 $14\arcmin \times14$\arcmin\ region, excluding the arms and shocks described above, is dominated by apparently
 isothermal fluctuations (bubbles) with a possible admixture (at the level of $\sim$30\%) of
 adiabatic (sound waves) and by isobaric structures. Larger features, of about 30 kpc, 
 show a stronger contribution from isobaric fluctuations. The results
 broadly agree with an AGN feedback model mediated by bubbles of
 relativistic plasma.
 \end{abstract}

\section{Introduction}
It is well known that the cores of relaxed clusters are strongly perturbed by the outflows of relativistic plasma from a central supermassive black hole \citep[e.g.,][]{bohringer95,churazov00,2000ApJ...534L.135M,fabian00,david01,nulsen02,birzan04,dunn06}, suggesting that the amount of energy supplied by the AGN is sufficient to offset gas cooling losses. In the vicinity of the AGN these outflows inflate ``cavities/bubbles'' in X-ray-emitting gas, which evolve under the action of a buoyancy force \cite[e.g.,][]{churazov01,2005MNRAS.357..242R}. Among plausible channels which transfer energy from the outflow to the gas two processes received much attention: shocks \cite[e.g,.][]{2003MNRAS.344L..43F,forman07,2011ApJ...726...86R} and purely subsonic interaction of the rising bubbles through the X-ray atmosphere, which uplifts cool gas, drives turbulence and excites gravity waves \cite[e.g.][]{2002MNRAS.332..729C}. When these processes are operating at the same time problems arise in differentiating among them in order to assess their relative roles. { In this paper we present a method to broadly identify the type of process, either adiabatic, isobaric or isothermal, that leads to the largest amplitude fluctuations}.

Here we use a simple technique (Churazov et al. in prep., see also \citealt{zhuravleva15}) of combining information contained in X-ray images in different energy bands to determine which process makes the largest contribution to the observed fluctuations of the intracluster medium (ICM) in M87/Virgo. We do this by calculating the cross power spectrum for two \chandra images and calculating the correlation coefficient and relative amplitudes of fluctuations in several energy bands, namely 0.5--1 keV, 1--3.5 keV and 3.5--7.5 keV. M87/Virgo is a canonical nearby example of a cool core cluster with prominent
substructure \citep[e.g.,][]{bohringer95,forman05,forman07,simionescu10,million10}. Throughout the paper
we adopt a distance to M87 of 16 Mpc, 1$'$ corresponds to 4.65
kpc. Its proximity and relatively low gas temperature makes it an
ideal object for study based on \chandra X-ray images. { In an accompanying paper we also apply this analysis to the Perseus cluster (Zhuravleva et al. in prep).}

In Sec. \ref{SB} we detail the effect that density fluctuations have on X-ray images in different bands and how this can be exploited to differentiate between isobaric, adiabatic and isothermal processes. In Sec. \ref{PS} we introduce the power and cross spectrum measures that will be used to characterize the images and we test these statistics with simulated cluster images in Sec. \ref{tests}. We proceed in Sec. \ref{data} with real cluster data, using \chandra images of M87 to determine the dominant type of fluctuation in different regions of the ICM. We interpret these results in term of the energy budget of the fluctuations in Sec. \ref{sec:discussion} and summarize our findings in Sec. \ref{sec:conclusion}.

\section{Energy dependent impact of fluctuations on X-ray images.}
\label{SB}
Many well known processes can cause fluctuations of the ICM density
and temperature, which in turn give rise to perturbed X-ray
images. Among them: (i) weak shocks and sound waves, (ii) bubbles of
relativistic plasma that make X-ray cavities in the gas, (iii) gas
displacement from its equilibrium position by buoyant bubbles and
subsonic gas motions\footnote{This list is by no means complete, but
  for the sake of simplicity we concentrate on these three
  processes.}. Each of these processes will lead to specific {\it
  energy dependent} signatures in the X-ray data. For instance, weak
shocks and sound waves cause density fluctuations that are positively
correlated with temperature fluctuations, as long as thermal
conduction is suppressed. On the other hand, cavities associated
with the bubbles of relativistic plasma, may have similar depressions
in X-ray images in all bands (if the X-ray emissivity of the plasma inside
the cavities is small). Finally, subsonically displaced gas lumps may
conserve their entropy (once again, provided that conduction is
suppressed) and remain in pressure equilibrium with the ambient gas. To
maintain pressure balance, these lumps will show anti-correlated
temperature and density fluctuations. Thus, even if the true ICM
equation of state is adiabatic with $\gamma=5/3$, the observed
correlations of the density and temperature fluctuations may not
correspond to the $\gamma=5/3$ equation of state. Below we classify
different types of fluctuations based on their {\it apparent}
equation of state, based on the relative amplitudes of fluctuations
present in X-ray images in different energy bands.

As mentioned above, the three aforementioned processes differ by the
temperature changes associated with a given amplitude of the density
fluctuation. Our goal is to use the images in different
energy bands to determine the type of process which gives rise to the
density fluctuations.  Assuming that the flux emitted per unit volume
is $F=\rho^2 \Lambda_b(T)$, where $\rho$ is the density and
$\Lambda_b(T)$ is the emissivity in band $b$ as a function of
temperature $T$, all that remains is to calculate the relation between
fluctuations of $\rho$ and $T$ for each type of process and the
emissivity $\Lambda_b(T)$ for each energy band.

We estimated $\Lambda_b(T)$ by folding an {\it XSpec} APEC model
through the response function of Chandra's ACIS-I detector, and
calculate the flux in each band as a function of temperature,
following the procedure described in \citet{forman07}. For this study
we use two energy bands: 1.0--3.5 keV and 3.5--7.5 keV (``soft'' and
``hard'' bands, respectively). The choice of the reference energy
bands is driven by two factors. On the one hand, using a combination
of images in very soft and very hard bands maximizes the sensitivity
to the temperature variations. On the other hand, we need sufficient
numbers of counts in each image to produce meaningful results. In
Churazov et al. (in prep) we address this issue and show that the pair we use
is close to optimal for the temperature characteristic of M87.

For unperturbed temperature and density $T_0$
and $\rho_0$, the perturbed values are 
\begin{enumerate}
\item $T=T_0(\rho/\rho_0)^{-1}$ for
an isobaric fluctuation, 
\item $T=T_0 (\rho/\rho_0)^{2/3}$ for adiabatic
fluctuations and 
\item $T=T_0$ for isothermal fluctuations. 
\end{enumerate}
Therefore, the change of the flux $\delta F_b$ in band
$b$ in response to a change in density 
\begin{equation}
\label{G}
\rho=G\times\rho_0
\end{equation}
 is
\begin{eqnarray}
\frac{\delta F_b}{F_b}&=\frac{\rho^2 \Lambda_b(T)-\rho_0^2 \Lambda_b(T_0)}{\rho_o^2 \Lambda_b(T_0)}=\frac{\rho_0^2 G^2 \Lambda_b(T_0G^\eta)}{\rho_0^2 \Lambda_b(T_0)} -1\\
&=\frac{G^2 \Lambda_b(T_0G^\eta)}{\Lambda_b(T_0)} -1
\end{eqnarray}
where the exponent $\eta$ is equal -1, 2/3 and 0 for isobaric,
adiabatic and isothermal processes, respectively. In the limit of
small density fluctuations $G=1+x$, where $x\ll 1$, the above
expression reduces to $\displaystyle \frac{\delta F_b}{F_b}=\left
(2+\eta \frac{d\log \Lambda_b}{d\log T} \right)x$.

\begin{figure}
\psfig{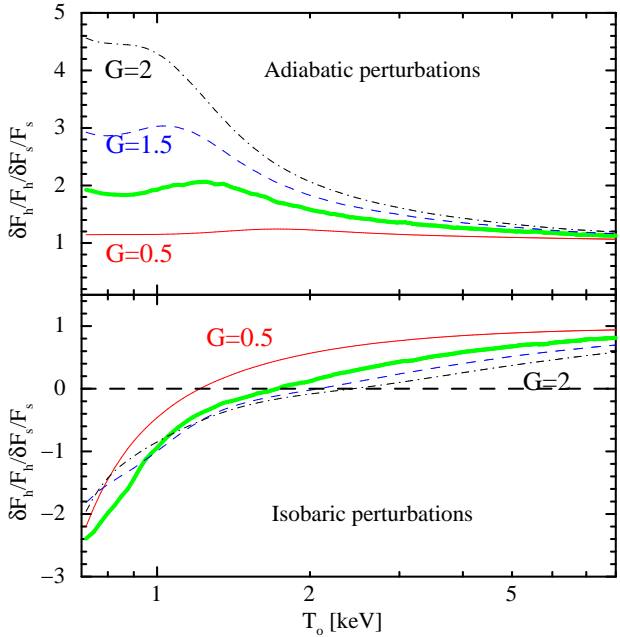}
\caption{
  Ratio of relative amplitudes of flux fluctuations in the ``hard''
  (3.5--7.5 keV) and ``soft'' (1.0--3.5 keV) bands as a function of
  initial temperature $T_0$, for adiabatic processes (top panel) and
  isobaric processes (bottom panel). A factor $G$ (
  see Eq.\ref{G}) characterizes the amplitude of the density
  fluctuation.  The thick green line in each plot corresponds to a
  limit small density fluctuations ($G-1\ll 1$), while the thin lines
  correspond to fluctuations with a given finite amplitude, as labeled
  in the plot. For the typical gas temperatures in M87 ($T\sim 2$ keV)
  the amplitudes for isobaric and adiabatic processes differ strongly,
  opening the possibility to use this difference to distinguish these
  processes in \chandra images.  }
\label{df}
\end{figure}

In Fig.\ref{df} we show the ratio of relative amplitudes $\frac{\delta
  F_h}{F_h}/\frac{\delta F_s}{F_s}$ of the fluctuations in the hard and soft band images 
corresponding to the same fluctuation in density. Adiabatic and
isobaric cases are shown in the top and bottom panels,
respectively. Isothermal fluctuations have equal amplitudes in all
bands, i.e. they produce a ratio equal to 1, irrespective of
temperature. The green thick line in each set corresponds to the limit
of small amplitude fluctuations. To illustrate the nonlinear effects
of large amplitude fluctuations we also show the curves for $G\equiv
\frac{\rho}{\rho_0}=0.5$ (red, solid line), $G=1.5$ (blue, dashed
line) and $G=2.0$ (black, dot-dashed line). In the temperature range
relevant to M87, 1.3--3 keV, the three types of process are clearly
distinguishable, producing larger, smaller or equal amplitude
fluctuations in the hard band than in the soft band for adiabatic,
isobaric and isothermal fluctuations, respectively. Factor of $\sim$2
fluctuations in density shows substantial deviations from the small
amplitude limit and marked asymmetry between positive and negative
fluctuations ($G=2$ or $G=0.5$). For smaller amplitudes (less than
$\sim$50\%), the green curves provides a reasonable approximation of the
flux ratio.

\begin{figure*}
\psfig{file=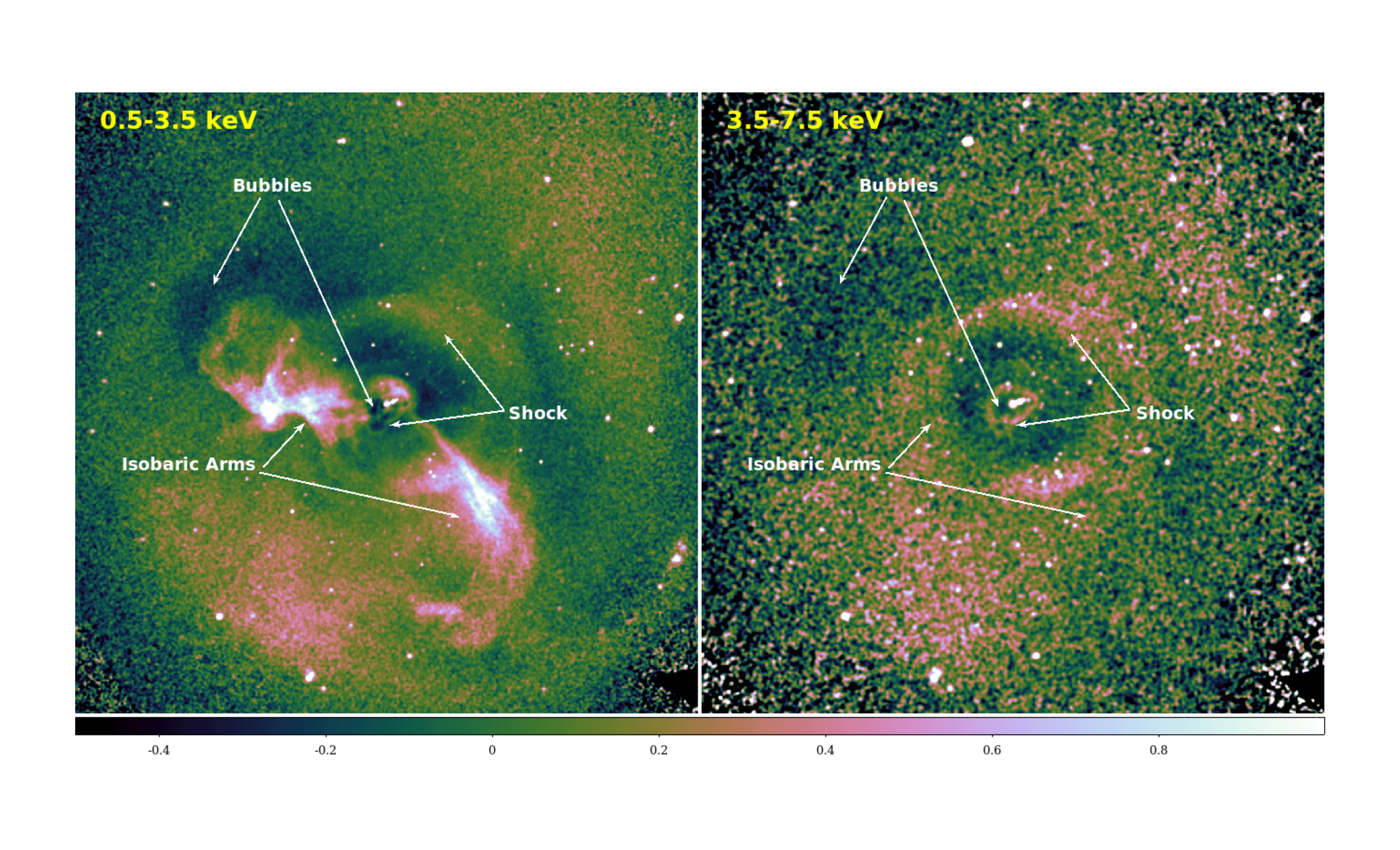,width=1.0\textwidth,angle=270}
\caption{0.5--3.5 keV and 3.5--7.5 keV images of M87 divided by their best-fitting $\beta$-models. The 0.5--3.5 keV image shows complicated substructure, while the 3.5--7.5 keV looks   strikingly different. As shown by \citet{forman07}, for gas  temperatures, characteristic for M87, the surface brightness in this  band is insensitive to isobaric fluctuations, but reflects projected pressure fluctuations. The lack of prominent ``arms''  in this image suggests that they are in pressure equilibrium with  the ambient gas. }
\label{images}
\end{figure*}

One noticeable feature of isobaric fluctuations is that around 2 keV,
which is close to the average temperature in M87, the ratio
$\frac{\delta F_h}{F_h}/ \frac{\delta F_s}{F_s}$ is close to
zero. This is because, for the hard (3.5-7.5 keV) band, $\displaystyle
\frac{d\log \Lambda}{d\log T} \sim 2$ at $T\sim 2$ keV and so
$\displaystyle \frac{\delta F_h}{F_h}\sim 0$. The 3.5--7.5 keV image
is therefore a good proxy for density fluctuations produced by
non-isobaric fluctuations \citep{forman07}, i.e., it is only sensitive
to adiabatic and isothermal fluctuations. In contrast, for the
1.0--3.5 keV band the emissivity is essentially independent of
temperature, as noted before. Thus $\displaystyle \frac{d\log
  \Lambda}{d\log T} \sim 0$ and $\frac{\delta F_s}{F_s} \sim 2
x$. Therefore this band is a proxy for density variations
independently of the type of process that produces them. Figure \ref{images} shows a soft (0.5--3.5 keV) and a hard (3.5--7.5 keV) image. The arms disappear in the hard band image, consistent with an isobaric nature interpretation, while the rings  appear as larger amplitude surface brightness fluctuations in the hard band than in the soft band image, as expected for adiabatic weak shocks. 

As is clear from Fig. \ref{df} the ratio $\frac{\delta F_h}{F_h}/
\frac{\delta F_s}{F_s}$ for isobaric fluctuations changes sign around
2 keV and is also sensitive to the amplitude of
fluctuations. Therefore one might expect to find either correlation or
anti-correlation of fluxes in these two bands, although the absolute
value of this ratio for isobaric fluctuations remains significantly
smaller than for isothermal or adiabatic fluctuations.

\section{Power spectra}
\label{PS}
We now proceed with the analysis of scale-dependent fluctuations,
including faint ones, which we would like to characterize in a
statistical sense through the power spectra and cross power spectra of two \chandra\ images. We will focus on fluctuations around the
large scale intensity profile of the cluster by first dividing the
images by the projected spherically symmetric $\beta$-model which best
fits the azimuthally-averaged radial profile. The fluctuation images $I=\frac{I_{raw}}{I_\beta}-1$ in each energy band are used in the analysis below. Here $I_{raw}$ is the original image and $I_\beta$ is the symmetric  $\beta$-model. The values of the slope $\beta$ and core radius $r_c$ of the $\beta$ models fitted to the image in each band are given in Sec. \ref{data}.

We will use the method described in \citet{rmsk} to filter the
fluctuation images in different spatial scales in order to calculate
the power and cross spectra. This method corrects for gaps and
irregular boundaries in the images. The 2D power spectrum can be
converted into a 3D power spectrum through the de-projection procedure
explained in \citet{churazov12} and finally converted into the
amplitude of fluctuations in 3D as $A_{3D}=\sqrt{P_{3D}*4\pi k^3/2}$,
where $k$ is the inverse of the spatial scale in units of
arcsec$^{-1}$, and $P_{3D}$ is the power per mode in the 3D power
spectrum { of emissivity fluctuations}, reconstructed from the observed 2D power spectrum. Note that we use $k=1/l$ where $l$ is the spatial scale, without the usual factor of $2\pi$ in this relation.  The
factor $2$ in the above expression is used to recast the amplitude in
terms of linear density fluctuations instead of volume emissivity
variations, since X-ray flux scales as $\propto \rho^2$ and for small
fluctuations this approximates to $\rho^2/\rho_o^2 = (1+x)^2 \sim
1+2x$ where $\rho_o$ is the unperturbed density and $x$ is the density
fluctuation. The cross spectra $P_{1,2}$ are calculated using a
similar procedure, by filtering the soft and hard band images with the
Mexican hat filter, multiplying the filtered images in two bands by
each other and calculating the mean value over all pixels.

In the discussion below, we assume that fluctuations, associated with
adiabatic, isobaric and isothermal processes are spatially
uncorrelated (in 3D). This assumption simplifies the interpretation of
the results by removing all cross-terms for different types of
fluctuations. While one can easily imagine a situations when
different types of processes are correlated (e.g., buoyant bubbles are
often accompanied by cool entrained gas), the procedure described below
should be able to establish which type of process dominates the
fluctuations in the images in a statistical sense.

\subsection{Correlation coefficient}

Since the relative amplitudes of the soft and hard band fluctuations
differ for different processes, fluctuations observed in two energy
bands are expected to perfectly match each other (i.e. $I_1=\alpha
I_2$, where $\alpha$ is a constant) only if the observed fluctuations
are dominated by one particular type of process. To this end, a handy
quantity is the correlation coefficient, which verifies if
fluctuations in one band are linearly related to the fluctuations in
another band. Given two fluctuation images, the correlation
coefficient $C_M$ can be computed as
\begin{equation}
\label{C}
 C_M=\frac{\sum I_1 \times I_2}{\sqrt{\sum I_1^2 \times \sum I_2^2}},
\end{equation}
where the subscript $M$ refers to the measured value as opposed to the
predicted value. In practice, we calculate a scale dependent
correlation coefficient $\displaystyle
C_M=\frac{P_{1,2}}{\sqrt{P_1P_2}}$, where $P_1$, $P_2$ and $P_{12}$
are the power spectra of $I_1$, $I_2$ and their cross spectrum,
respectively. The contribution of the Poisson noise to $P_1$, $P_2$ is
removed by generating several images of Poisson fluctuations around the fitted $\beta$-model, calculating their power spectra and averaging the results. The averaged Poisson noise power spectrum is then subtracted from $P_1$ and $P_2$.  Unlike the standard definition of the correlation
coefficient in signal processing, we keep the sign of correlation
coefficient, i.e. the allowed range of $C_M$ is between -1 and 1, to differentiate correlated from anti-correlated fluctuations.

For a single type of process with small amplitude fluctuations in an
isothermal cluster, the image in one band is just a linear
transformation of the image in the other. Therefore, the correlation
coefficient should be 1 or -1. In reality, clusters might be perturbed by different processes with
comparable amplitudes. Some of these will have larger responses in the
hard band than in the soft band and some other, spatially uncorrelated
fluctuations, will have the opposite effect, so that the final images
in both bands might also appear uncorrelated. This mixture of
processes will be reflected in the measured correlation coefficient.

For a combination of different (independent) processes, the
fluctuation image in band $b$ can be expressed as $I_b\propto \sum
\alpha_{b,i}X_i$, where $X_i$ is proportional to the line-of-sight
emission measure fluctuation relative to the unperturbed model,
associated with the process $i$, and $\alpha_{b,i}=\left (2+\eta_i
\frac{d\log \Lambda_b}{d\log T} \right)$ is the response in band $b$
to this fluctuation. The expected correlation coefficient between
soft and hard band images is then:
\begin{eqnarray}
C_E&= \frac{\sum_i \alpha_{1,i} \alpha_{2,i} \langle X_i^2\rangle}{\sqrt{\sum_i \alpha_{1,i}\langle X_i^2\rangle \times \sum_j \alpha_{2,j}\langle X_j^2\rangle}},
\label{eq:C}
\end{eqnarray}
where the cross terms $\langle X_i X_j\rangle$ have been dropped by
the assumption of uncorrelated variations. It is clear from this
equation that if a single process dominates the density fluctuations,
e.g. $\langle X_i^2\rangle>>\langle X_j^2\rangle$ for all $j$ and all
$\alpha_{1,i}$ and $\alpha_{2,i}$ are of the same order, then the
correlation coefficient converges to $\displaystyle
C_E=\frac{\alpha_{1,i} \alpha_{2,i}}{|\alpha_{1,i} \alpha_{2,i}|}$,
i.e., to a value of 1 or -1. On the other hand, for a mixture of
processes with comparable amplitudes, the correlation coefficient  takes a value $|C_E| < 1$, that can be predicted once the values of
$\alpha$ are known.

Notice, that $C_M$ will be poorly constrained if the leading process
does not produce a strong signal in one of the bands (i.e. if the response in
$I_1$ or $I_2$ to this process is close to zero). For $T=2$ keV gas,
this is the case for isobaric processes in the 3.5--7.5 keV band (see
Fig. \ref{df}). As is clear from this figure, even for small variations
of temperature around 2 keV, the sign of the ratio of amplitudes
changes, leading to large uncertainty in the value of the correlation coefficient.
 
\subsection{Relative amplitudes of soft and hard band fluctuations }
Another useful quantity is the relative amplitudes of fluctuations in
two energy bands $R$.  The correlation coefficient $C$ defined above
in Eq. \ref{C} equals the cross spectrum normalized by the variance of
both bands. To estimate the relative amplitude of fluctuations, $R$,
we normalize the cross spectrum by the variance of the soft band only,
\begin{equation}
\label{eq:R}
R= \frac{P_{1,2}}{P_1}=C\times \frac{\sqrt{P_2 }}{\sqrt{ P_1}}.
\end{equation}
$R$ is proportional to the correlation coefficient and to the ratio of the amplitudes of the fluctuations in the hard and soft bands. The fact that we multiply the simple ratio of powers by the correlation coefficient preferentially highlights the ratio of correlated (or anti-correlated) fluctuations, which are the ones revealing we wish to study. We use the form
$\frac{P_{1,2}}{P_1}$ to calculate this quantity rather than $C\times
\frac{\sqrt{P_2 }}{\sqrt{ P_1}}$ because the first expression removes
much of the uncertainty related to subtracting Poisson noise, which is
high in the hard band. Since this noise is uncorrelated in the two bands,
the Poisson noise cancels out in the cross term $P_{1,2}$, and the denominator only
includes the soft band power $P_1$ which has a lower Poisson noise
level.

\subsection{$R-C$ map}

In the case of two or three types of independent fluctuation
processes being comparably important, we expect that the relative
amplitudes and the correlation coefficient of the soft and hard band
images will have intermediate values.

\begin{figure}
\psfig{file=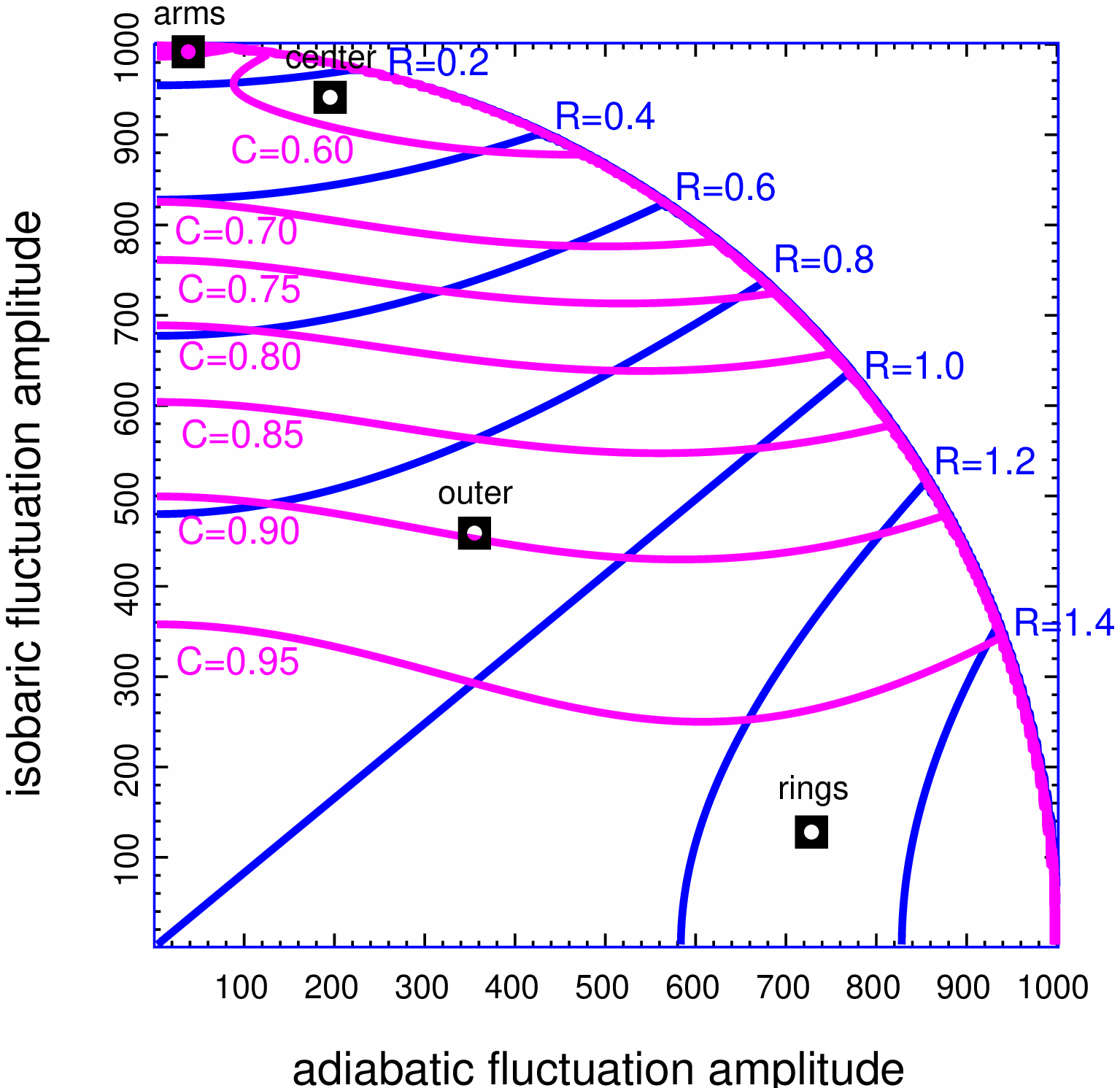,width=0.5\textwidth}
\psfig{file=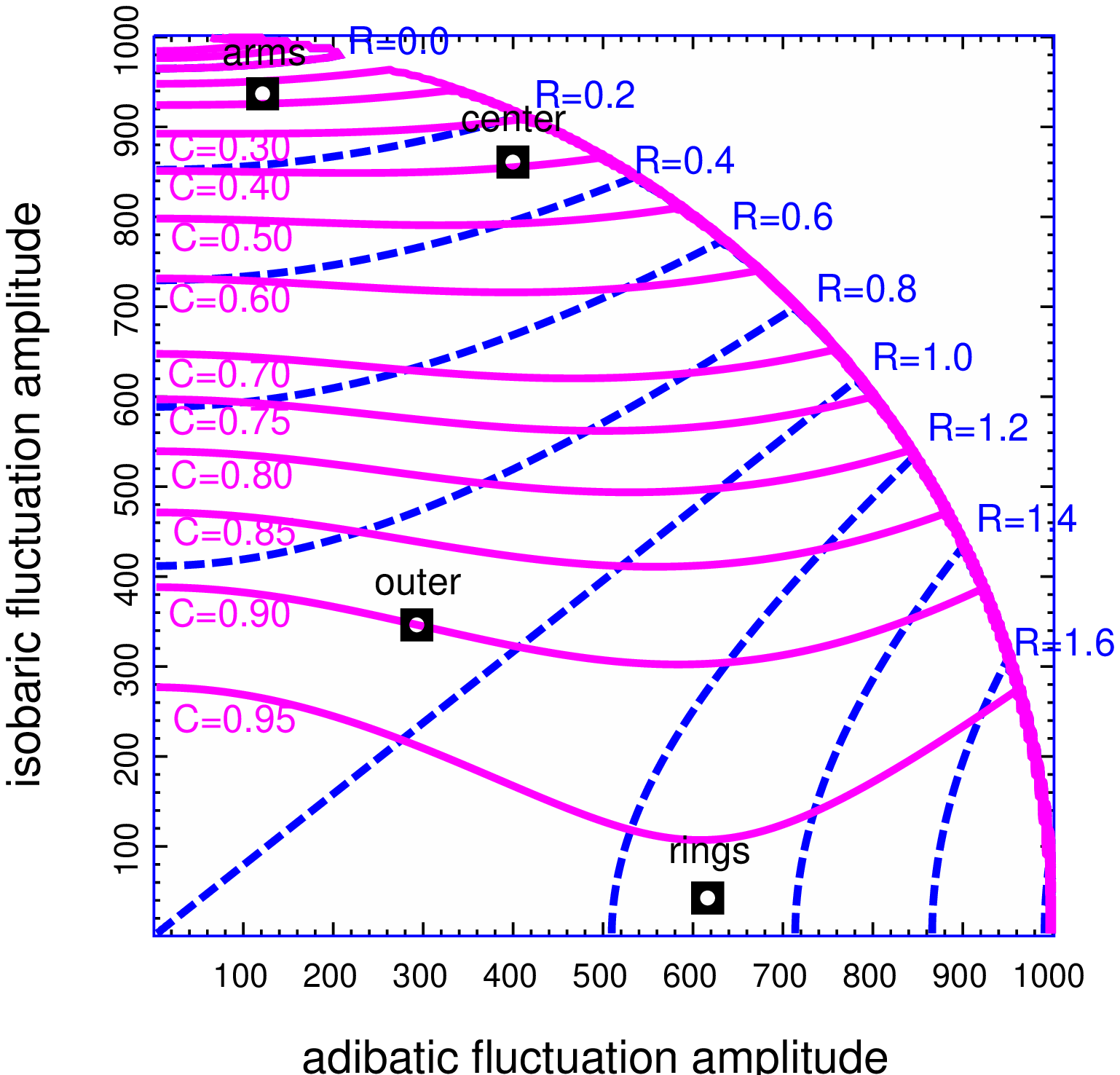,width=0.48\textwidth}
\caption{R-C maps: Predicted correlation coefficient $C$ (magenta
  contours) and $R$ (blue contours) for a mixture of adiabatic,
  isobaric and isothermal processes, calculated assuming
  small-amplitude uncorrelated fluctuations of the three
  processes. The total variance of all three processes is normalised to
  unity.  The amplitude of adiabatic fluctuations increases towards
  the right, isobaric fluctuations upwards and isothermal fluctuations
  decrease with increasing distance from the origin. The plot on the
  top shows the predicted values for an unperturbed temperature $T_0=
  2$ keV and the plot on the bottom assumes $T_0=1.6$ keV. These two temperatures bracket the range of temperatures observed for M87 in the radial range $0.5-5\arcmin$. Boxes
  schematically show the typical values of $R$ and $C$ calculated for
  several regions in M87. The ``arms'' and the ``center'' regions are
  clearly dominated by isobaric fluctuations. On the contrary, the
  circular ``ring'' is dominated by adiabatic fluctuations. The outer
  region (see Sec. \ref{outer} below) is consistent with a mixture of several
  processes, whith isothermal fluctuations making the largest
  contribution.}
\label{coh_ratio}
\end{figure}

Figure \ref{coh_ratio} shows the expected correlation coefficient
(magenta contours) and amplitude ratio (blue contours) for the simulated images in
the 1--3.5 keV and 3.5--7.5 keV bands. The top figure is calculated
for an unperturbed temperature $T_0= 2$ keV and the bottom figure is
for $T_0=1.6$ keV. We normalize the total variance of density
fluctuations, i.e., the sum of the variances corresponding to isobaric,
adiabatic and isothermal processes, to 1. Therefore, in 3D Cartesian
space, where the amplitude of each type of fluctuation corresponds
to one of three orthogonal axes, the parameter space considered here
corresponds to one octant of a spherical shell, which we project
into the (x,y) plane. The amplitude of adiabatic fluctuations
increases towards the right, isobaric fluctuations upwards and
isothermal fluctuations decrease with increasing distance to the
origin $z=\sqrt{1-(x^2+y^2)}$, so that pure adiabatic fluctuations are
mapped into the bottom right corner, pure isobaric in the top left and
pure isothermal to the origin. The correlation coefficient and
amplitude ratio contours form an almost orthogonal patterns in these
$R-C$ maps. Therefore, for our idealised model of three uncorrelated
processes, it is possible to find the coordinates of the intersection
of the relevant $R$ and $C$ contours and from these to determine the
fraction of the total variance produced by each type of density
fluctuation. These unperturbed temperatures, $T_0= 1.6$ and $T_0= 2$
keV were chosen to bracket the average temperature in M87 within
6\arcmin\ from the centre, which is the area that will be studied
below. For the real cluster, or simulations calculated using the real
cluster temperature radial profile, the measured $R$ and $C$
coordinates should lie between the limiting maps shown in Fig. \ref{coh_ratio}.

\section{Testing $R$ and $C$ with simulated images}
\label{tests}
\begin{figure}
\psfig{file=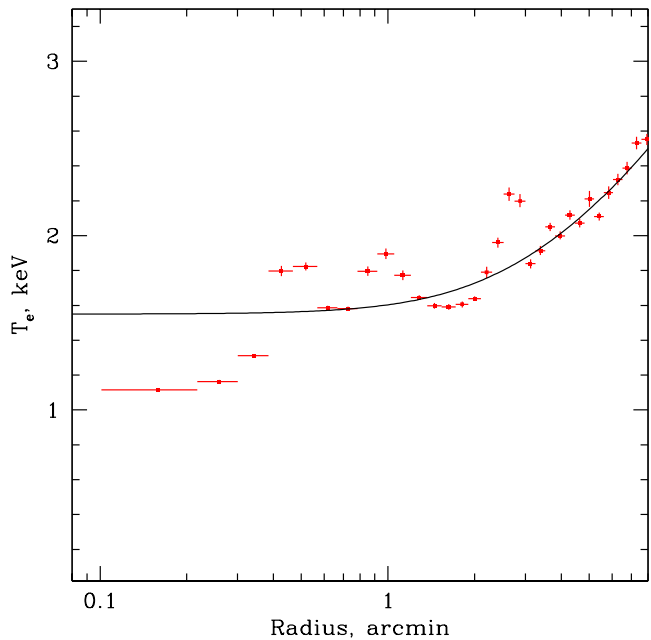,width=0.5\textwidth}
\caption{Azimuthally-averaged radial profile for the deprojected temperature in M87 from \citet{churazov08}.}
\label{fig:tprof}
\end{figure}

Even in the simplest scenario, when all fluctuations in the ICM
correspond to only one type of process, other complications remain, e.g., radial variations of the mean temperature profile, projection effects and,
possibly, nonlinear responses to large-amplitude fluctuations. Even though the
temperature profile is known, the location of each individual
fluctuation along the line of sight is not. The latter effect can be corrected  
for only in a statistical sense.  We proceed by testing the
effect of radial temperature profiles and different amplitudes of
fluctuation on $R$ and $C$ with simulated cluster images using the
de-projected radial temperature profile for M87 from \citet{churazov08},
shown in Fig. \ref{fig:tprof}.

The images are produced by generating three independent realizations
of the density fluctuation patterns (one for each of the three
processes) as $\rho_i/\rho_0=e^{\delta_i}$, where $\delta_i$ is a
Gaussian random field with a given 3D power spectrum $P$, representing
fluctuations for process $i$. This log-normal form is used to allow
for relatively large amplitude density fluctuations. The
temperature fluctuations $T_i/T_0$ are calculated for each process
separately, e.g. $T_i/T_0=\left(\rho_i/\rho_0 \right)^{2/3}$ for an 
adiabatic process. These values are then used to calculate the
fluctuation of the volume emissivity
$e_i/e_0=\frac{\rho_i^2\Lambda_b(T_i)}{\rho_0^2\Lambda(T_0)}$. The
final value of the emissivity in each band is calculated as $e=e_0
\sum_{i=1}^3\frac{e_i}{e_0}$.  The resulting 3D cubes are projected
onto one plane to produce simulated images in each band. For these
tests we assume, for each process, the same shape of the power spectrum
$P\propto k^{-3}$, where $k$ is the wavenumber. The
choice of a $k^{-3}$ spectrum is rather arbitrary. It has the advantage
that in 3D it corresponds to the same amplitude at all scales and is
not far from the canonical Kolmogorov slope of $k^{-11/3}$, which is consistent with the observed spectrum in M87 and Perseus \citep{zhuravleva14b}.

The soft (1.0-3.5 keV) and hard (3.5-7.5 keV) band simulated images
are processed in the same way as real images (see \S\ref{data} below),
except for the correction for Poisson noise. Namely, we fit $\beta$-model
to each image, divide the image by the $\beta$-model and calculate $C$
and $R$ over a range of angular scales.

In the following two subsections (\S \ref{sec:adiabatic} and \S
\ref{sec:isobaric}) we perform this analysis for pure adiabatic or
pure isobaric fluctuations, varying their amplitudes and assuming
either an isothermal cluster or one with a radial temperature
profile. Isothermal fluctuations are not considered separately, since
they have the same impact on the emissivity in all bands, so the
emissivity patterns and images are identical in all bands, except for
the overall radial temperature profile which largely cancels out
by fitting and dividing the images by the projected $\beta$-model. In
\S~\ref{sec:mix} we consider a mixture of all three processes.

\subsection{Pure adiabatic fluctuations}
\label{sec:adiabatic}
\begin{figure*}
\plotfour{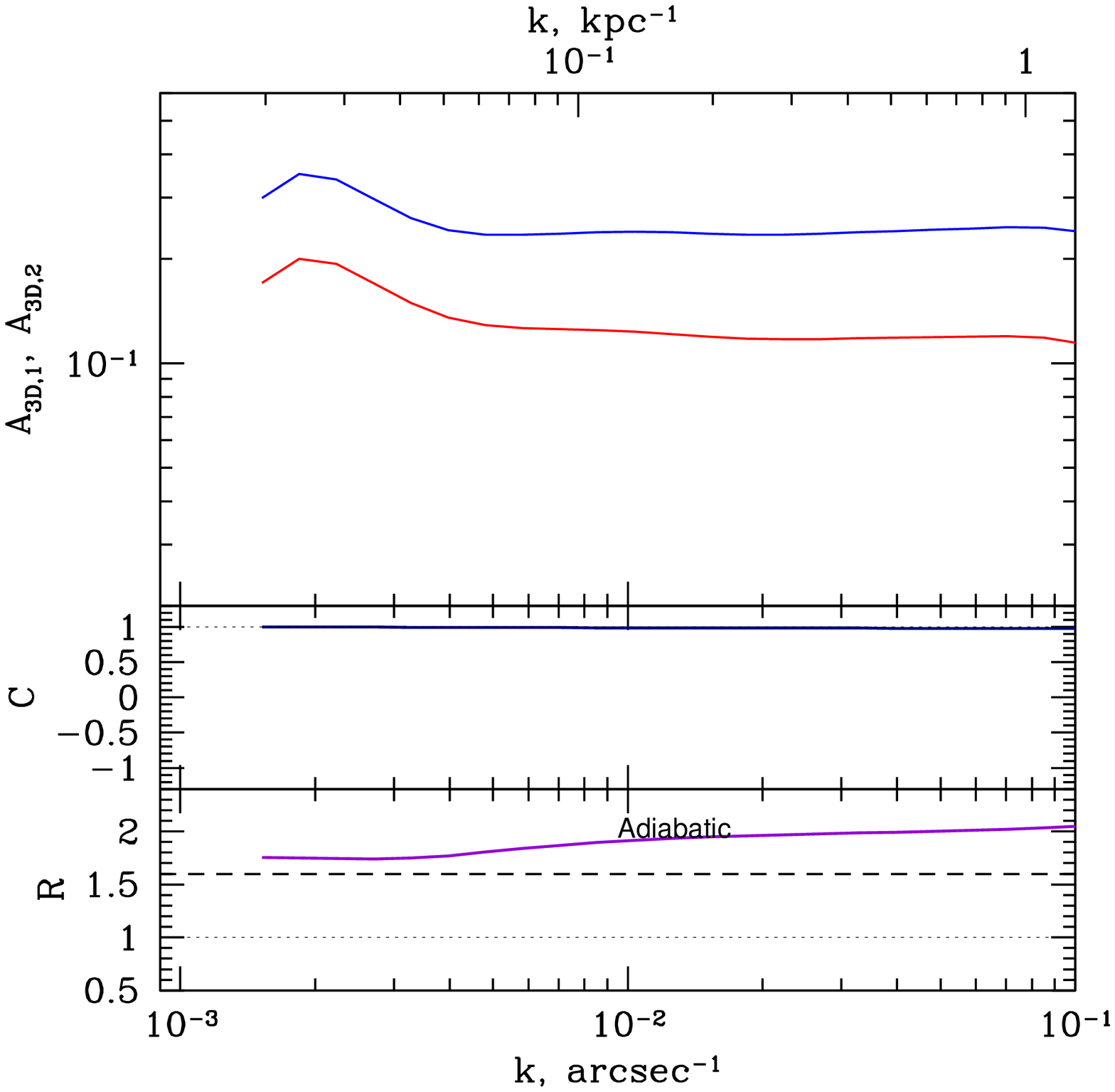}{acc30A9_B0_T0_T_logn.ps}{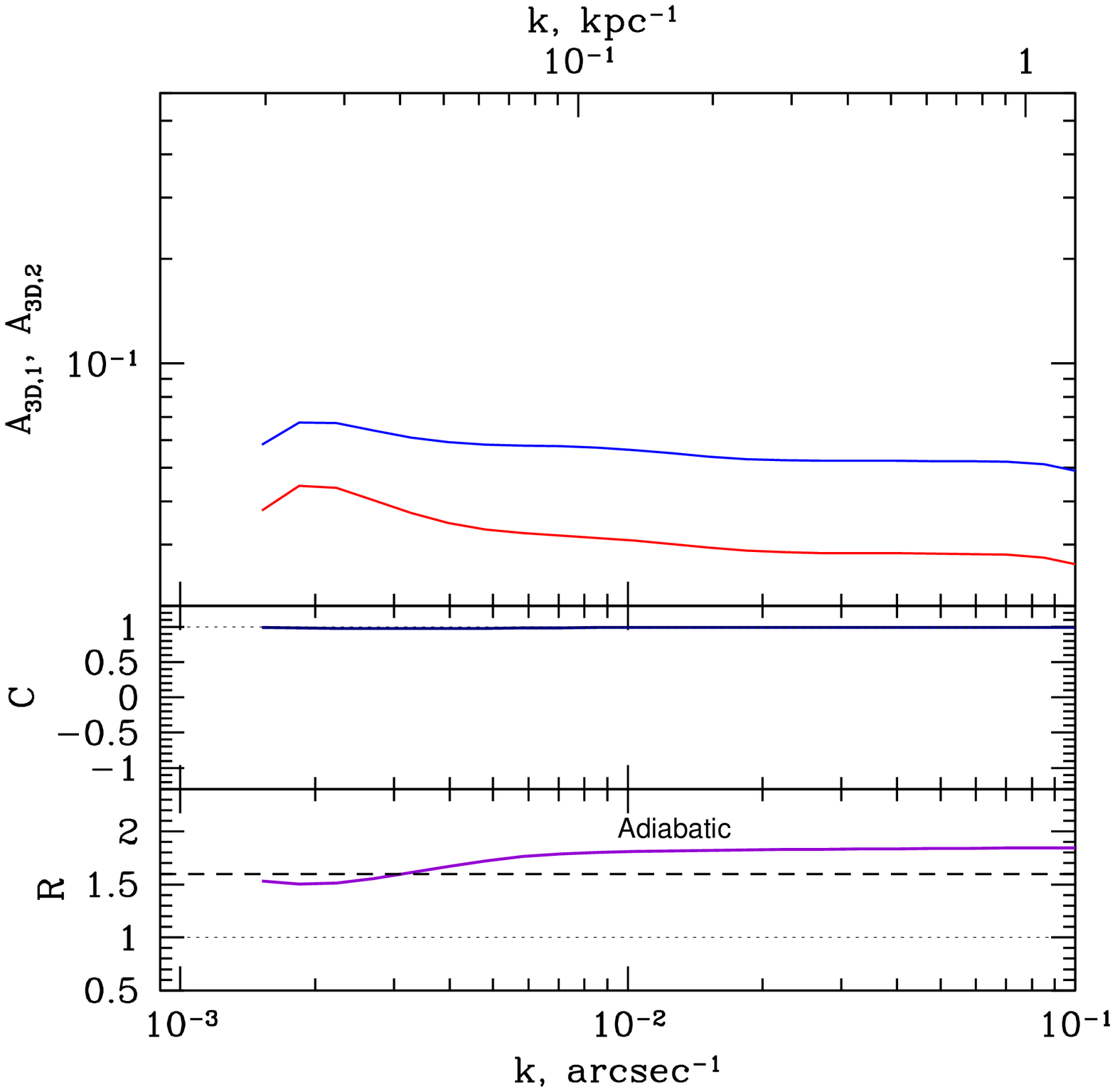}{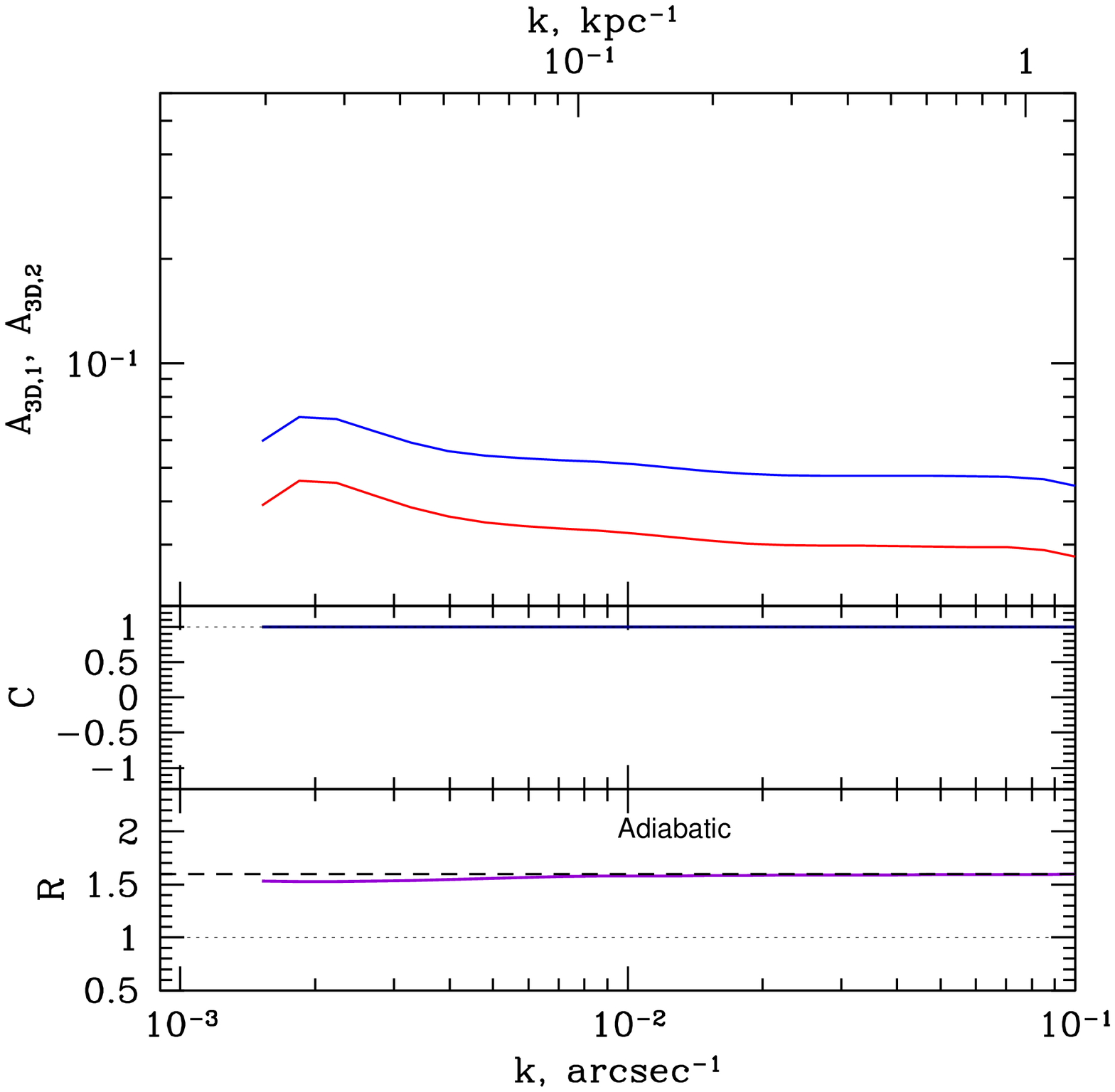}

\caption{Test of the recovered $C$ and $R$ values for simulated images
  with pure adiabatic fluctuations. Left: Average radial
  temperature profile as measured for M87. Right: isothermal
  cluster at $T=2$ keV. The top panels correspond to large amplitude fluctuations and the lower panels to lower amplitudes. The amplitude of the fluctuations observed in M87 has
  intermediate values. In the top
  section of each panel, the blue and red lines show the hard and soft
  amplitude spectra, respectively. In the middle section, solid blue lines 
  show the correlation coefficient $C$ and in the bottom section the solid  
 purple lines show the relative amplitude $R$. Pure adiabatic fluctuations
  show high correlation coefficient $C$ and high value of $R>1$.  The dashed lines in the bottom panel mark the expected value of $R$ for adiabatic fluctuations around $T_0=2$ keV.}
\label{adiabatic}
\end{figure*}

Figure \ref{adiabatic} summarizes the results of simulations for pure
adiabatic fluctuations. Left and right columns correspond to a cluster
with a radial temperature profile and to an isothermal cluster,
respectively. Top and bottom rows are for large and small
amplitudes\footnote{Since our simulated box has finite size, the
  projected surface brightness distribution does not correspond to a
  perfect $\beta$-model. This causes a hump in the amplitude on the
  largest scales, comparable to the size of the box.},
respectively. The top section in each plot shows the spectra of the
recovered amplitudes in the soft (red) and hard (blue) bands. In the
middle sections we plot the correlation coefficient $C$, and the pink
line in the bottom panel shows the ratio $R$ as defined in
eq. \ref{eq:R}.  The dashed line shows the expected value of $R$ for
an adiabatic process at an unperturbed temperature $T_0= 2$ keV.

For adiabatic fluctuations we can conclude that, for these
temperatures and energy bands, the correlation coefficient $C$ is in
the range 0.95--1, and $R$ is between 1.5 and 2, even for a varying
temperature profile and for large or small fluctuations. Large
fluctuations introduce a trend in the value of $R$, where hard band
fluctuations increasingly dominate towards smaller spatial scales, but
their value remains in the range expected for an adiabatic
process. Therefore, pure adiabatic fluctuations should be clearly
identifiable in the data by their very high correlation coefficient
and large amplitude ratio, $C\sim 1, R>1$.

For a temperature of 2 keV the expected value of $R$ for adiabatic
fluctuations is $\sim 1.6$ (see horizontal dashed line in
Fig.~\ref{adiabatic}). Note that for the temperature profile adopted
here (see Fig.~\ref{fig:tprof}), the temperature is about 1.7 keV over a 
significant part of the simulated volume. The green curves in the top
panel in Fig. \ref{df} show that for this temperature, the expected
ratio of amplitudes is closer to 1.8, in excellent agreement with the
$R$ value of these simulations. This explains why the recovered value
of $R$ is larger in the left column of Fig.~\ref{adiabatic}, which uses a more realistic temperature profile.

\subsection{Pure isobaric fluctuations}
\label{sec:isobaric}
The situation is more complex for isobaric fluctuations. As can be
seen in Fig. \ref{df}, the hard band response is close to zero in the
temperature range of interest, i.e. isobaric density fluctuations
might show up clearly in the soft band but be weak or absent in the
hard band. Additionally, the response in the hard band changes from positive to negative
values as the temperature fluctuates around $\sim 2$ keV and this
critical temperature also depends on the amplitude of the
fluctuations. Therefore, the hard band response depends sensitively on
the amplitude of the fluctuations and on the local unperturbed
temperature, which can produce correlated, anti-correlated or
apparently unrelated fluctuations in the soft and hard band images.

\begin{figure*}
\plotfour{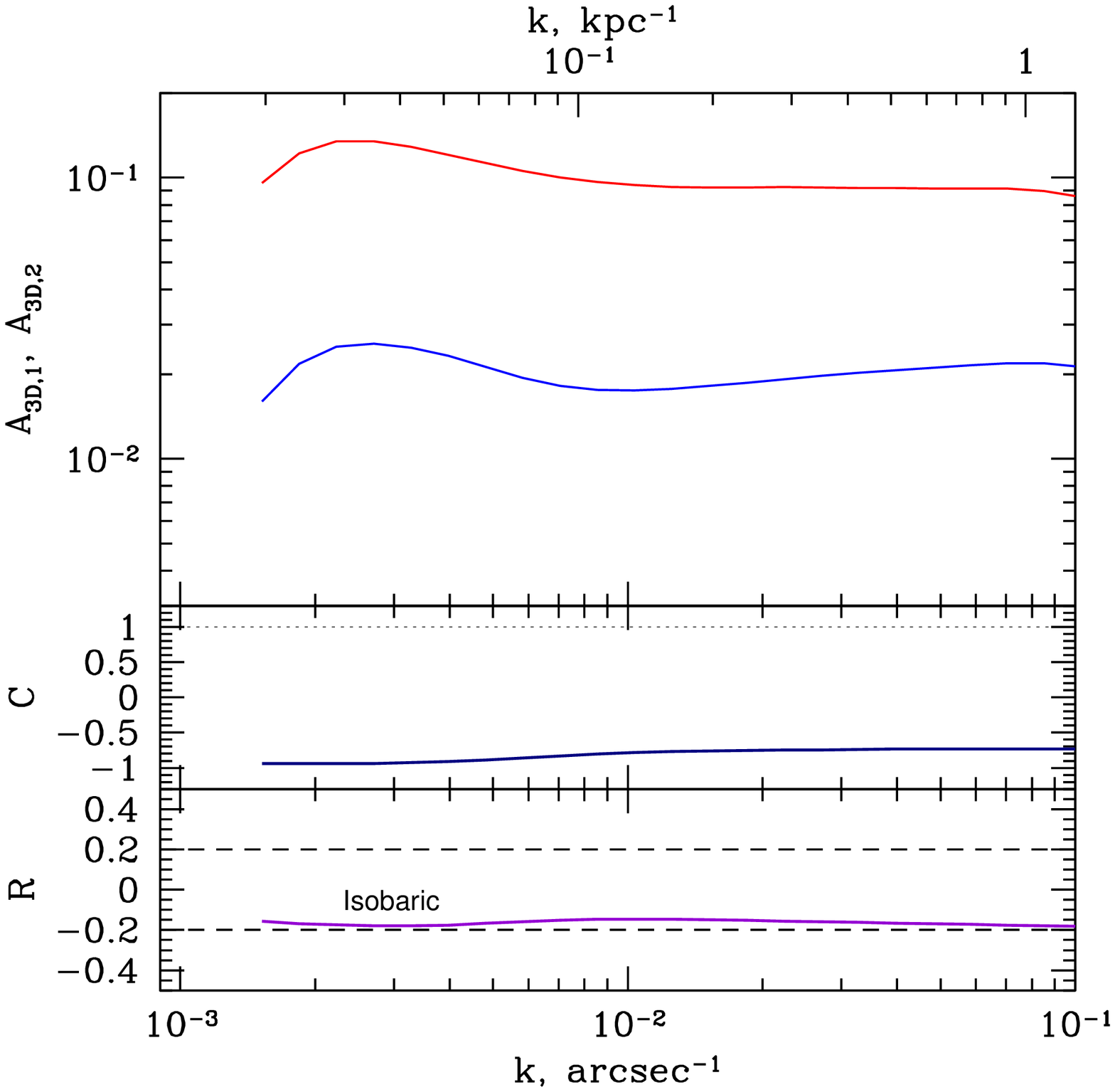}{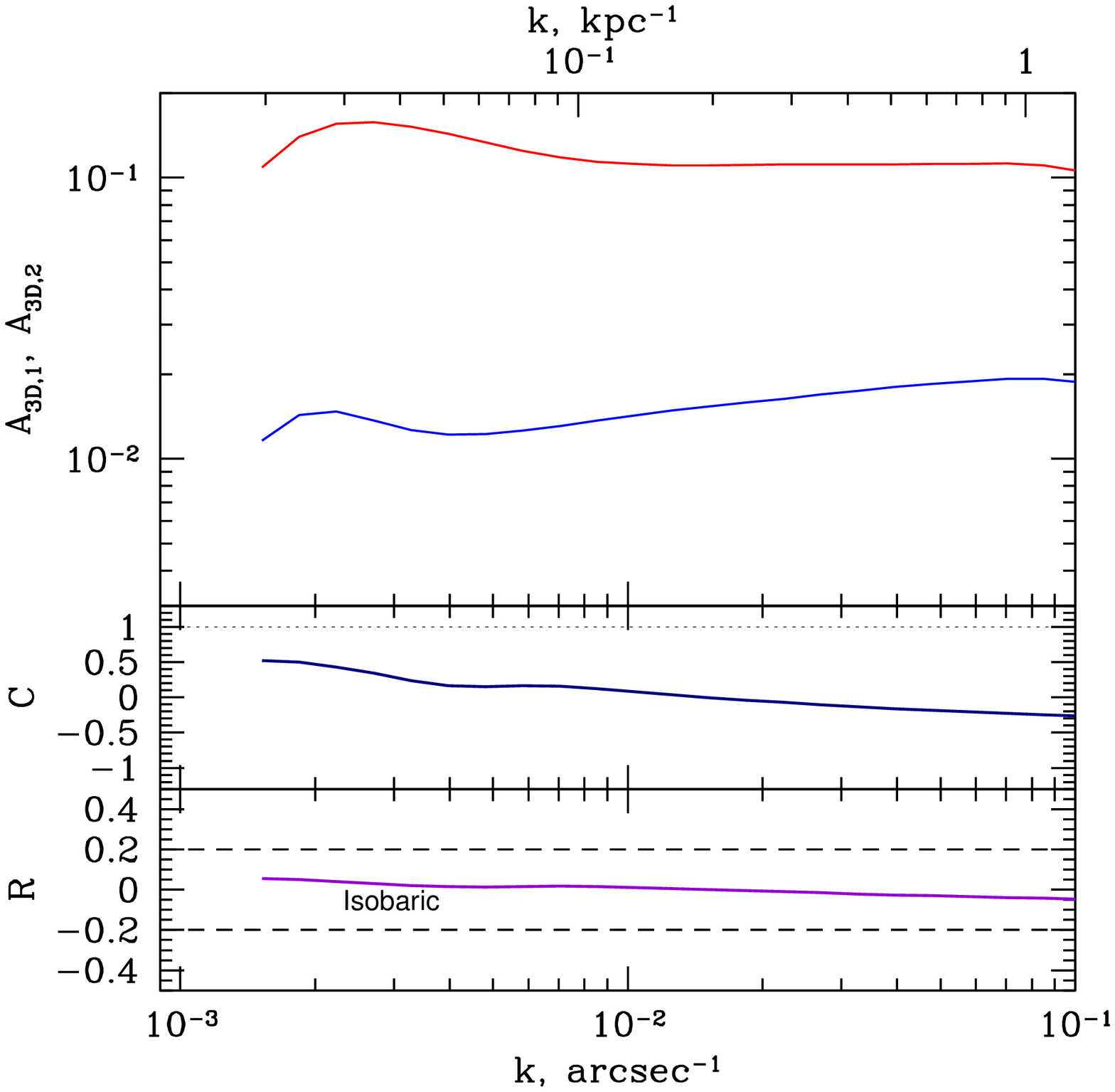}{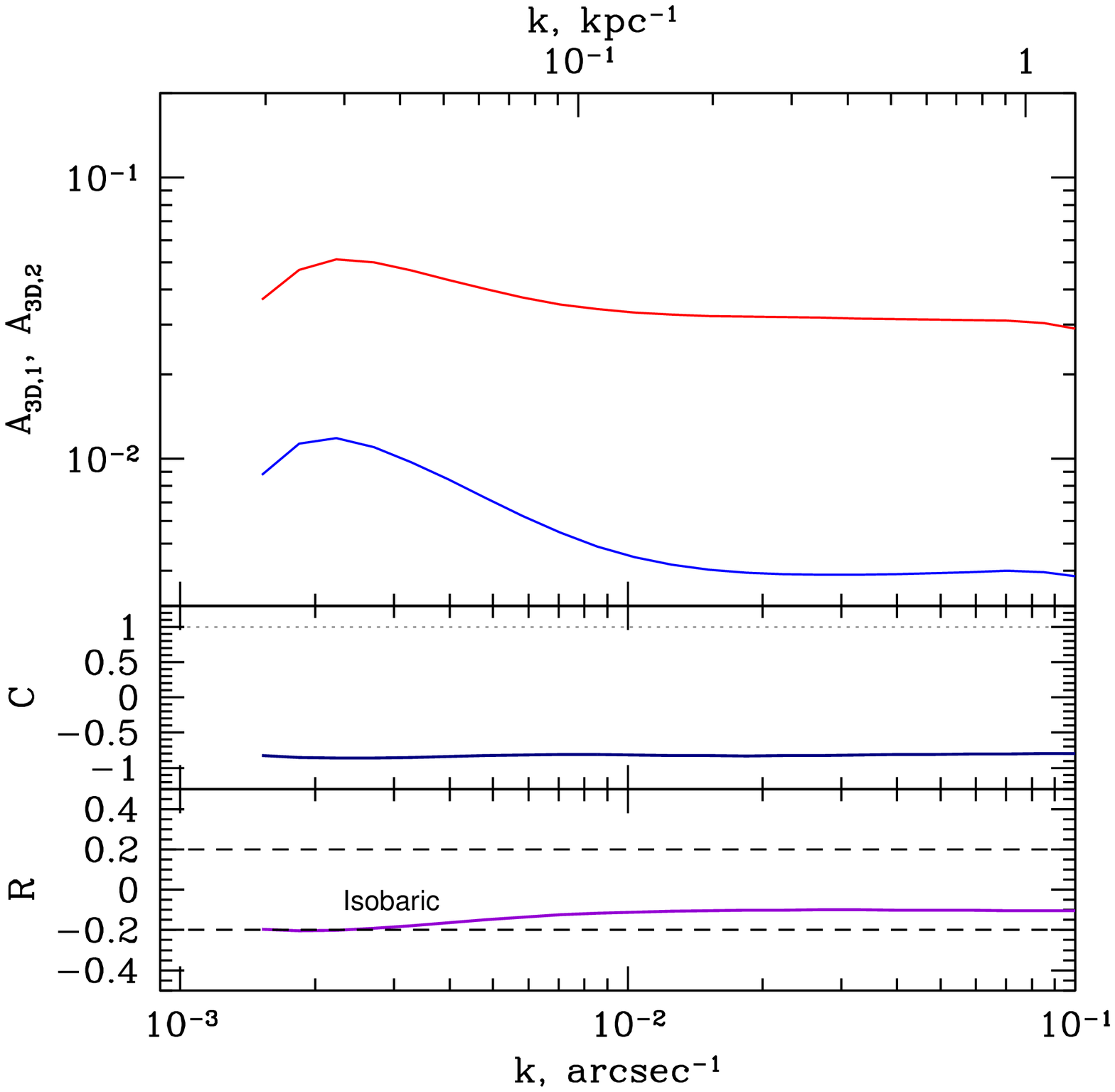}{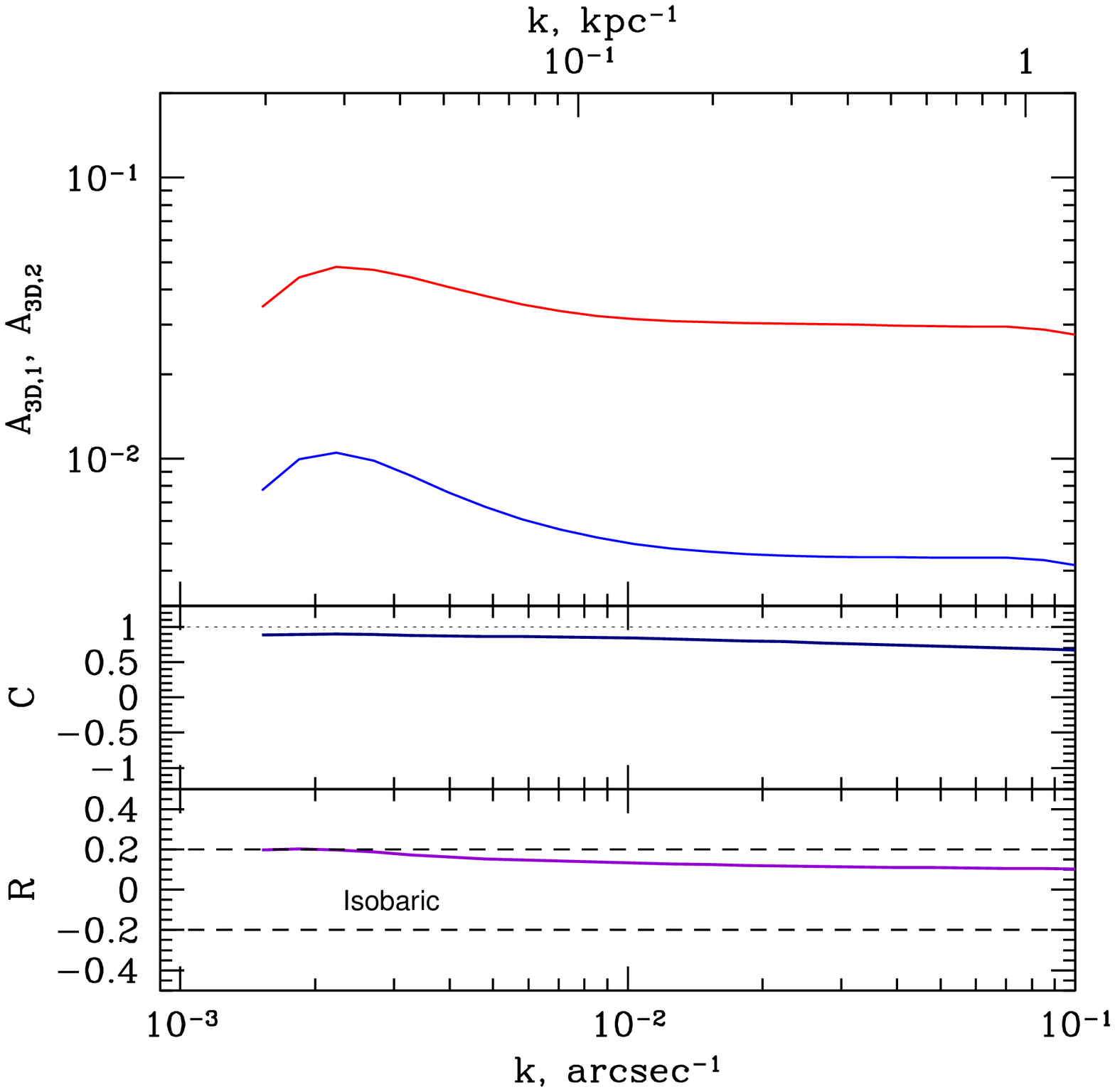}
\caption{The same as in Fig.~\ref{adiabatic} but for pure isobaric
  fluctuations.  As expected, pure isobaric fluctuations
  lead to low values of $R<1$ and small (or even negative) correlation coefficient $C$
  (except for the case of a perfectly isothermal cluster and very
  small fluctuations). The dashed lines in the bottom panel mark the range of values of $R$ expected for isobaric fluctuations around $T_0=2keV$.}
\label{isobaric}
\end{figure*}

Fig.~\ref{isobaric} shows the same simulations as in
Fig.~\ref{adiabatic} but for pure isobaric fluctuations.  In all cases,
the hard band amplitudes are smaller by a factor of a few than in the
soft band, and the amplitude ratio is low, as expected.  Very small
fluctuations around $T_0=2$ keV (bottom right panel) remain in the
positive response range and the correlation coefficient remains high
and positive. Large amplitude fluctuations on the other hand, produce
a negative hard band response, so the fluctuations can be highly
anti-correlated. For the left panels, which are simulated using the
radial temperature profile of Fig. \ref{fig:tprof}, the majority of
the simulated region is within 6\arcmin\ from the centre and therefore
has a temperature lower than 2 keV, producing mostly anti-correlated
hard and soft fluctuations.

Isobaric fluctuations can be identified more reliably through the low
amplitude of the hard band fluctuations ($|R|<<1$), while the
correlation coefficient must be interpreted with care, considering the
amplitude of the fluctuations and the local unperturbed temperature,
although small ($C \sim 0$) and negative values of $C$ always
correspond to isobaric fluctuations.

\subsection{Mixture of three processes}
\label{sec:mix}
We now consider a combination of three different types of processes by
combining the simulated images described above and including
`isothermal" simulated images as well, where the fluctuation
amplitudes are the same in soft and hard bands. To construct the
images we simply generate three patterns of density fluctuations
(modulated by a global $\beta$-model), convert each pattern into soft
and hard band emissivities, add the flux contributions from the three
processes in each volume element and project onto one plane.  We
measure the correlation coefficient $C$ and the ratio $R$ of each pair
of images and compare these values with the $R-C$ maps, derived for an
isothermal cluster and linear fluctuations. As before we use
simulated images in the 1--3.5 keV and 3.5--7.5 keV bands and vary the
relative amplitudes of the three processes in independent runs.

\begin{figure}
\psfig{file=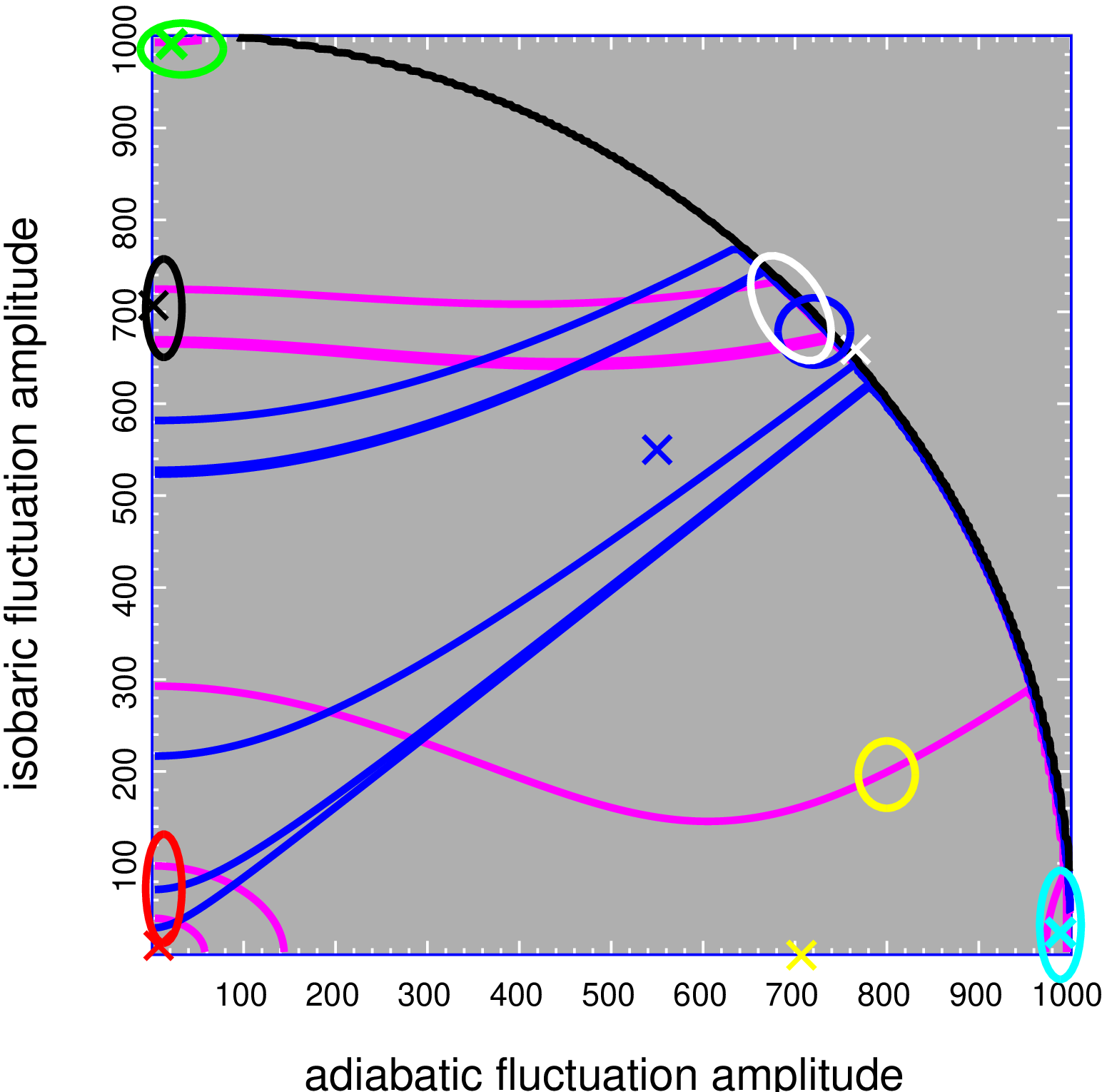,width=0.48\textwidth}
\caption{Simulated and recovered relative amplitudes of adiabatic,
  isobaric and isothermal fluctuations for 7 pairs of simulated
  images. Input values are marked with crosses. Blue and magenta lines
  show the contours, corresponding to $C$ and $R$ value obtained from
  the simulated images. The intersection of these contours define the recovered
  values of relative amplitudes of adiabatic, isobaric and isothermal
  fluctuations. These intersections are shown with ellipses (same color
  as the input values). When a single process dominates the density
  fluctuations, the recovered values are close to the input
  values. For a combination of two or more processes the accuracy
  deteriorates but the dominating processes and their relative
  fractions are qualitatively recovered. }
\label{intersectionsT2}
\end{figure}

The relative contributions of the three processes in the simulated
images were chosen to cover several limiting cases: each single
process dominating, equal contributions of two processes and equal
contributions from the three processes. The positions corresponding to 
these setups are marked with crosses in Fig.~\ref{intersectionsT2},
with single processes marked in red, green and cyan. Two
equal-amplitude processes cases are marked in yellow, black and white, and the case where the three processes have equal amplitudes is marked in blue. We then processed each pair
of simulated images as we would do with the real data. We calculated
$R(k_{ref})$ and $C(k_{ref})$ values for each pair and plotted the
resulting values as contours in blue and magenta in
Fig. \ref{intersectionsT2}. Here we choose $k_{ref}=10^{-1}~$kpc,
corresponding to the middle of the range of scales we probe in M87. The intersection of the
contours yields the recovered values of relative amplitudes of the
adiabatic, isothermal and isobaric process. These intersections are
marked with ellipses in the same color as the corresponding input parameters that are marked with
crosses.

The $R-C$ map, that we use to compare the outcome of the simulations,
was calculated for small-amplitude fluctuations around a constant
temperature $T_0=1.6$ keV, which is representative of the unperturbed
temperature range used in the simulations. The results show that, even
in the presence of a global temperature gradient, the leading type of
fluctuation is correctly recovered by comparing the measured values of $C$ and
$R$ with the predictions for a constant temperature cluster. Indeed, for
the majority of cases the recovered characteristics of fluctuations (ellipses
in Fig.~\ref{intersectionsT2}) are close to the input values (crosses
in Fig.~\ref{intersectionsT2}). For a combination of two or more
processes the accuracy deteriorates but the dominating processes and
their relative fractions can still be qualitatively recovered.

We note that the relative responses of both bands are sensitive to the
amplitude of fluctuations when the unperturbed temperature is low (see
low $T_0$ range in Fig. \ref{df}). In this case large amplitude
fluctuations can lead to $R$ values that fall outside the range of the
$R-C$ maps in Fig. \ref{coh_ratio} that was calculated assuming 
small-amplitude fluctuations. Large-amplitude adiabatic fluctuations
will simply produce larger $R$ values than those predicted for the
pure adiabatic case (right corner of the $R-C$ map), while large isobaric
fluctuations can have $R$ values smaller than the minimum predicted for small-amplitude fluctuations 
(top left corner of the map). Nevertheless, even for large-amplitude
fluctuations, adiabatic and isobaric fluctuations remain
well-separated in the $R-C$ map.

\section{Nature of the surface brightness fluctuations in M87}
\label{data}
We use archival \chandra\ observations of M87 with ObsIDs 2707, 3717,
5826, 5827, 5828, 6186, 7210, 7211, 7212 and 11783. The data were
processed following the procedure described in \cite{vikhlinin05}. The
co-added final image has a pixel size of 1\arcsec\ and fairly uniform
total exposure of about 420 ks within 7\arcmin\ in radius of the
centre, which is the area we will use for the following analysis.  The
well studied X-ray jet \citep[e.g.,][]{marshall02,harris03} and the
point sources, mainly corresponding to X-ray binaries
\citep{jordan04}, contaminate the signal from the ICM fluctuations and
were excised from the images before further analysis.

We use X-ray images of M87 in the 1--3.5 keV and 3.5--7.5 keV bands
to calculate the power and cross-power spectra of fluctuations, the correlation
coefficient $C$ and the amplitude ratio $R$. We fit $\beta$-models
to the images in each energy band and then divide the images by these models to
remove the large scale symmetric structure. The best fitting
$\beta$-model parameters are $R_c=0.23 \arcmin$ and $\beta=0.37$ for
the 1--3.5 keV image and $R_c=0.03\arcmin$ and $\beta=0.30$ for the
3.5--7.5 keV image. The smaller value of $\beta$ for the hard band
reflects the global radial temperature increase that makes the hard
band image less peaked. The resulting ``flattened'' images are shown
 in Fig. \ref{images}.

\begin{figure}
\psfig{file=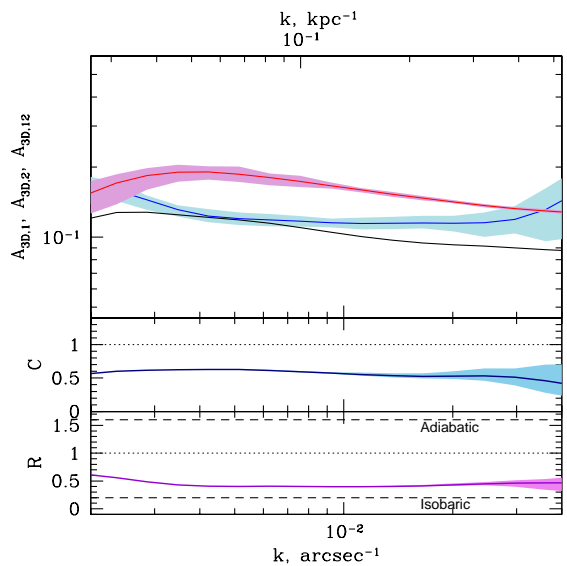,width=0.45\textwidth}
\caption{Power and cross-power spectra, correlation coefficient and
  ratio $R$ for the $14\arcmin \times 14\arcmin$ Chandra images of M87
  in the 1--3.5 keV and 3.5--7.5 keV bands. {\bf Top panel}: power
  spectra of fluctuation images in the soft (red) and hard (blue)
  bands. Shaded regions show the uncertainty in the measured
  spectra. The cross-power spectrum between these bands is shown with
  the black solid line. {\bf Middle panel}: correlation coefficient
  $C$. {\bf Bottom panel:} amplitudes ratio $R=P_{12}/P_{1}$. The dashed lines mark the expected value of $C$ for pure  isobaric and adiabatic fluctuations, around $T_0=2$ keV.  The low values of
  $R$ and $C$ indicate that isobaric fluctuations account for 70--80\%
  of the variance in this region. In the soft band the total variance
  is dominated by the contribution of filaments/arms of cool gas. }
\label{mask1}
\end{figure}

We first calculate the amplitude and cross spectrum for the entire
region, masking out only point sources and the jet. This region of the
ICM is perturbed by different processes so our aim is to establish
whether one of these clearly dominates the fluctuations in the
cluster.  The resulting spectra are shown in Fig.~\ref{mask1}. The
amplitude spectra in the top panel show a very slight dependence of
amplitude with spatial scale. Perfectly horizontal amplitude spectra
correspond to a 3D power spectrum $P_{3D}\propto k^{-3}$ so this
slight decrease of amplitude with $k$ indicates a slightly steeper
$P_{3D}$. This is consistent with the power spectral slope obtained for M87 by \citet{zhuravleva14b}, although in that paper the arms structure was excluded from the analysis making the amplitudes smaller, and for the Perseus cluster by
\citet{zhuravleva15}. The small $R$ and $C$ values reveal the strong
contribution of isobaric processes to the fluctuation pattern of this
region. For spatial scales of about 100\arcsec\ we find $R=0.35$ and
$C=0.5$. This value of $R$ occurs only at the top left corner of the
$R-C$ maps in Fig. \ref{coh_ratio}, where isobaric fluctuations are
responsible for about 70--80\% of the total variance. In this corner, 
the coherence is very sensitive to the underlying temperature so we only
remark that, as expected for both maps, the absolute value of the
coherence should be significantly lower than 1 but its precise value
is hard to predict.


We now proceed with the analysis of selected regions of the image. We
identified structures which might be dominated by different
processes. From Fig. \ref{images} it
is clear that the ring structures around 3\arcmin\ from the centre
correspond to adiabatic fluctuations \citep[see e.g.,][where this
  structure is identified as a weak shock]{forman07}, while the arms
stretching to the east and south-west are isobaric, since they do not
appear in the hard band image. Guided by these arguments we subdivided
the image into three regions: ``arms'', ``ring'' and ``outer''.  We
now extract the power spectra of these regions to verify if their
$R-C$ values match our expectations and establish the contribution of
the different processes in regions with less prominent features.

\subsection{Arms}
The filamentary regions extending to about 7\arcmin\ from the centre
towards the east and south-west, the arms, were identified in the
X-ray image of M87 by \citet{bohringer95} who pointed out their
spatial correlation with similar structures seen in the radio band. In
\citet{churazov01}, these structures were explained as gas entrained
from the core region by rising buoyant bubbles of relativistic plasma.
Later \citet{molendi02} noted that these regions were significantly
colder than the surrounding medium, and required a two-temperature
model for the cluster, with temperatures as low as 0.8--1 keV for the
arms. \citet{forman05} produced new temperature maps of the M87
cluster confirming the lower temperature of both arms while finding no
strong difference in the metallicity of the gas in the arms compared to the
surrounding medium.
 
\begin{figure*}
\psfig{file=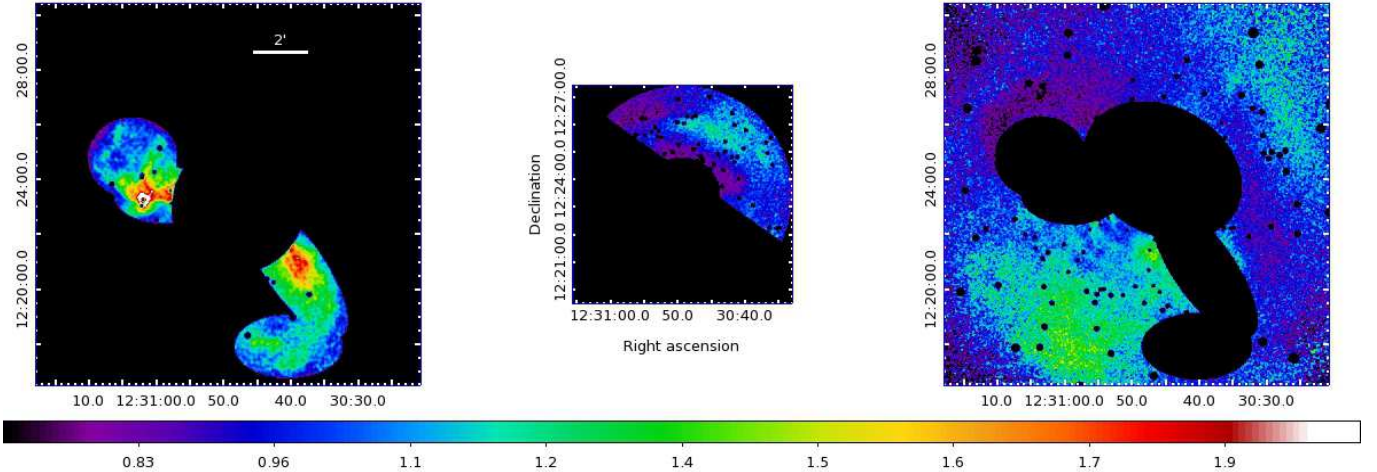,width=1.0\textwidth}
\caption{Selected image regions. Left: arms, center: shock, right: rest. The sections correspond to the same 1--3.5 keV image of M87, divided by the best-fitting $\beta$-model. All images are on the same scale. The size of the largest image is 14\arcmin  $\times$ 14\arcmin. }
\label{regions}
\end{figure*}

\begin{figure*}
\plotthree{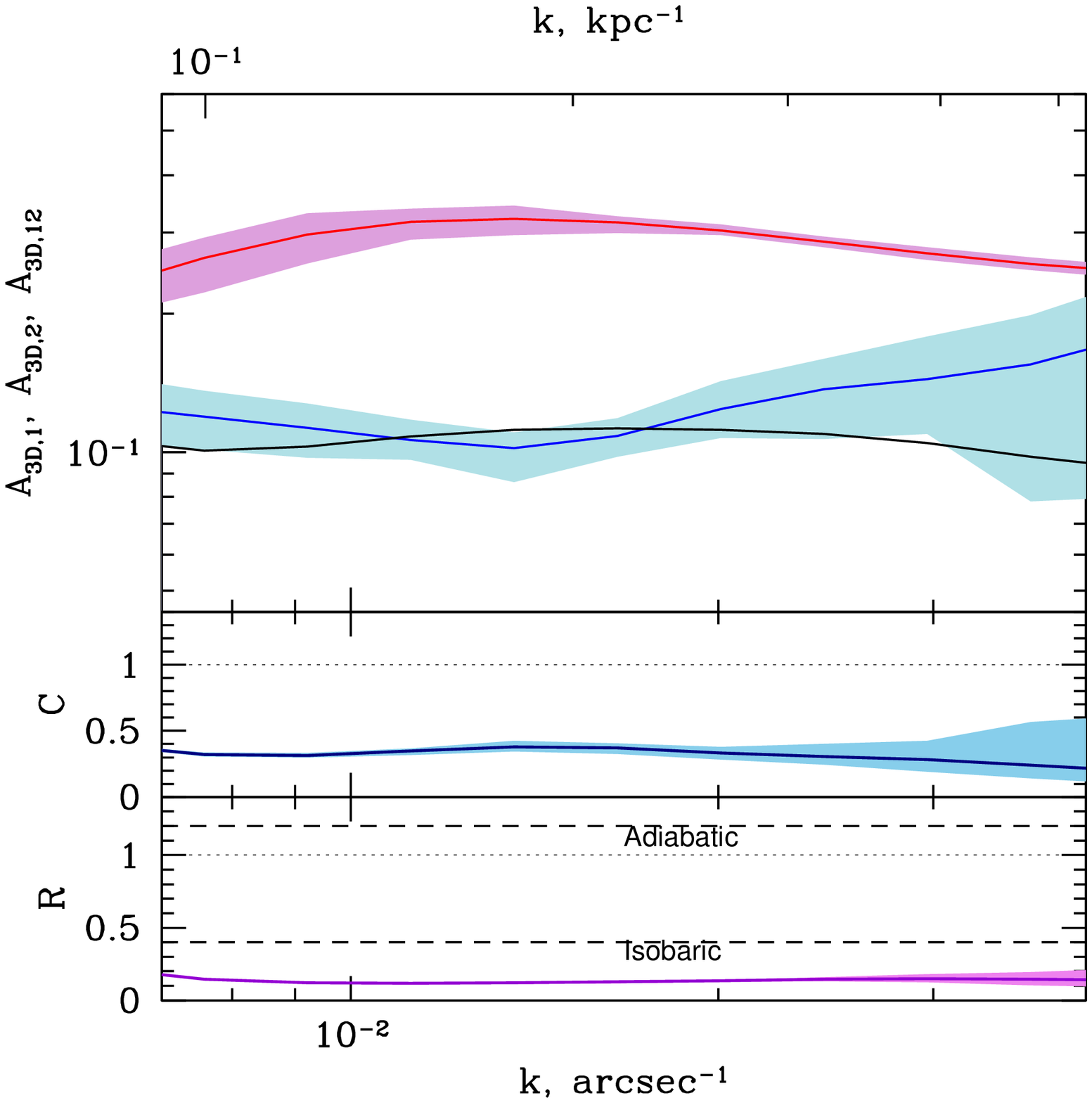}{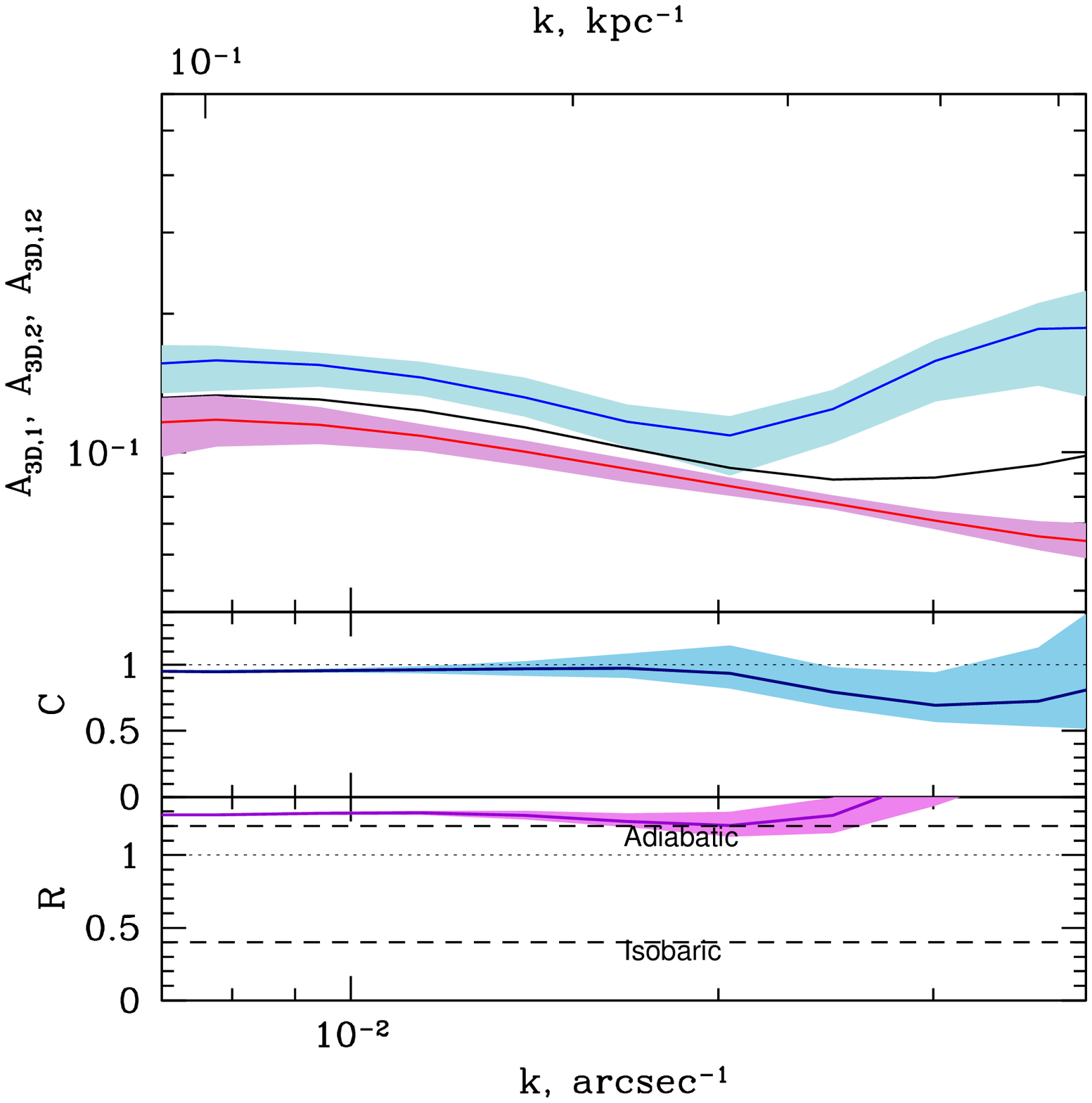}{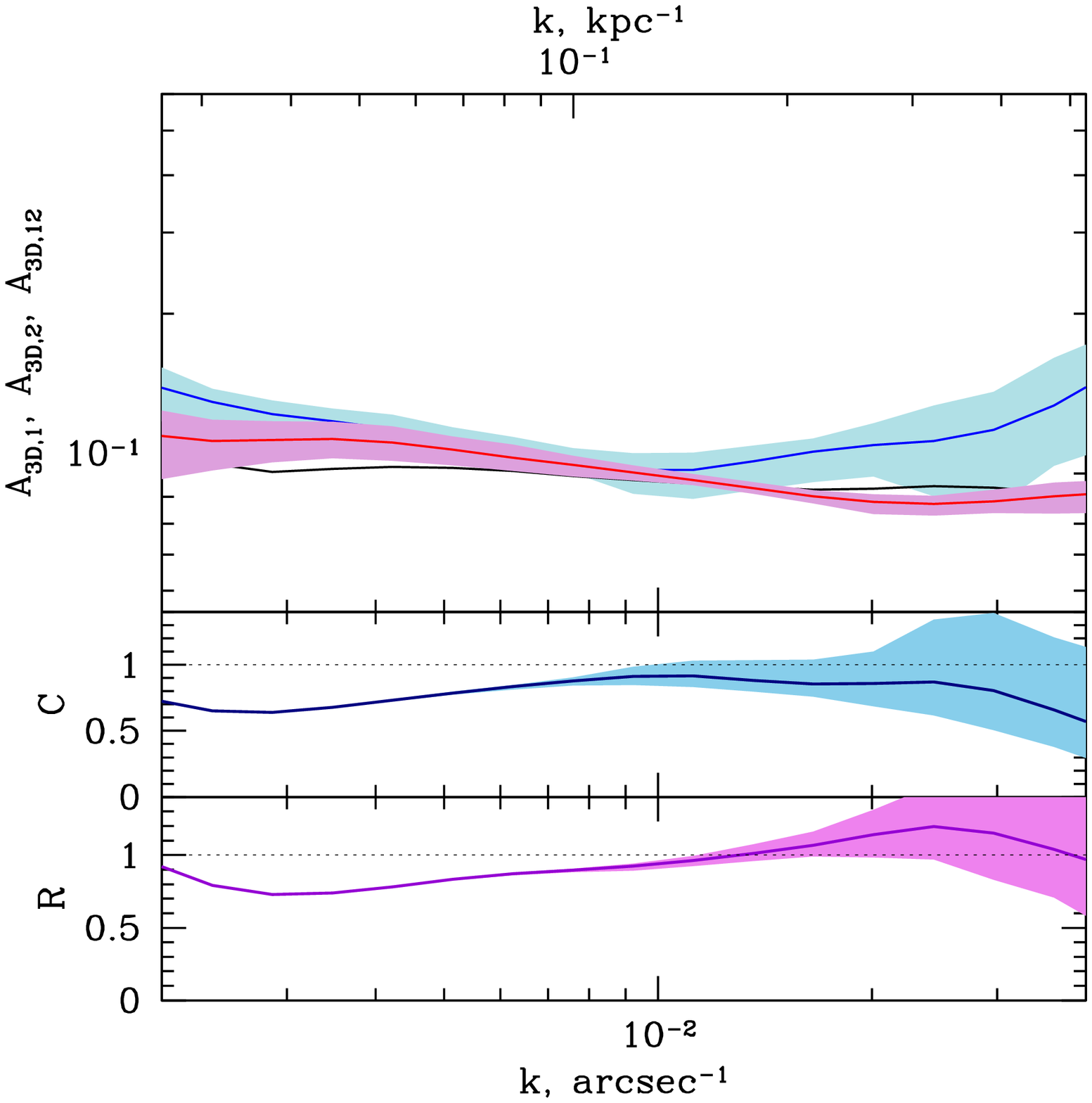}
\caption{The same as in Fig.~\ref{mask1} for three separate regions
  shown in Fig.~\ref{regions}. {\bf Left panel:} the region dominated
  by cool arms. The value of $R$ is close to that expected for
  isobaric fluctuations around mean temperature $T_0= 2$ keV. The
  correlation coefficient is small, as expected. {\bf Central panel:}
  the shock region.  Two dashed black lines correspond to pure
  adiabatic and isobaric processes for 3 keV gas. The values of $R$
  and $C$ suggest that adiabatic process (weak shock) dominates. {\bf
    Rights panel:} "outer" region (no arms, no central region, no
  shock) is only robust for $k\la 10^{-2}$arcsec$^{-1}$ since there
  are very few counts in this region and the hard band power drops by
  an order of magnitude below the Poisson noise level. Around and
  below $k=10^{-2}$~arcsec$^{-1}$ both the correlation coefficient and
  ratio are below unity (see text for details). }
\label{sections}
\end{figure*}

The part of the M87 image selected for our analysis of the arms is
shown in the left panel of Fig.~\ref{regions}. The width of the
selected region is about 130\arcsec , so we will restrict the analysis
to wave numbers $k>8\times10^{-3}$~arcsec$^{-1}$. The power spectra are
shown in the left panel in Fig.~\ref{sections}. On the smallest
scales, with $k>2\times10^{-2}$~arcsec$^{-1}$, unresolved point sources
make a significant contribution to the variance of the hard band
image. Since the spectra of unresolved sources are typically much
harder than the spectra of the M87 gas, these sources create an upturn
in the hard band power spectrum and also affect the amplitude ratio and
correlation coefficient. Since we are interested in the arm structures only, we restrict the
analysis to $k<2\times 10^{-2}$~arcsec$^{-1}$. In this range of scales,
the correlation coefficient and amplitude ratios are small with values
of $R\sim 0.1$ and $C\sim 0.35$, very similar to those of the entire
image, so the same interpretation applies here, only shifting the
relative importance of the isobaric fluctuations to even higher
values. The corresponding region in the $R-C$ maps is labeled `arms"
in Fig. \ref{coh_ratio}.

The value of $R\sim 0.2$ can be compared with the predictions for
isobaric fluctuations plotted in the bottom panel in
Fig. \ref{df}. For an initial temperature of $T_0 \sim 2$ keV, small
variations of density (green curve) will produce an amplitude ratio of
this order of magnitude. Pure isothermal ($R\sim1$) and adiabatic
processes ($R>1$) can be safely excluded, while isobaric fluctuations
in these regions are completely consistent with the amplitude ratios.

\subsection{Rings at 2\arcmin--3\arcmin.75 (11--17 kpc)}

The ring sections of increased surface brightness at radii of
3--3.75\arcmin\ from the centre were first identified by
\citet{young02} from \chandra\ images. \citet{forman05} present a
temperature map of this region showing an increase in temperature at
the rings, compared to the immediate surrounding gas, which together
with the nearly circular appearance of the ring support its
interpretation as a weak shock \citet[see also][]{forman07}. \citet{million10} used deep \chandra\ images to
confirm its weak shock nature.

The ring appears more prominent in the 3.5--7.5 keV band than in the
softer bands, consistent with an adiabatic origin (top panel in
Fig. \ref{df}).  We calculated $R$ and $C$ for a section of the image
covering the northern part of the ring, excising only the point sources
from this region, as shown in the central panel in Fig. \ref{regions}.
The thickness of the section including the ring is about 150\arcsec
, so we will focus on size scales smaller or equal to this, with $k
>6\times 10^{-3}$arcsec$^{-1}$. As can be seen in the central panel in
Fig. \ref{sections}, around $k=10^{-2}$arcsec$^{-1}, C\ga0.95$ and
$R=1.3$. The temperature in this region is about 2 keV, so we should
interpret the results using the top panel in
Fig. \ref{coh_ratio}. The location of the intersections of the corresponding
contours suggests the following distribution of amplitudes of the
contributing processes: adiabatic $\sim$ 0.8, isobaric $\sim$ 0.3,
isothermal $\sim$ 0.5. In terms of variance $\sim$ 60\% of the total
variance can be attributed to adiabatic processes, $\sim$10\% of the
variance to isobaric processes and $\sim$30\% of the variance to isothermal
processes. For higher mean temperatures, the adiabatic contribution
would increase further. Therefore, adiabatic fluctuations dominate the
variance in this region although there can be a non-negligible
contribution from at least one other process.

\subsection{Outer region}
\label{outer}
We now proceed with the analysis of the outer region shown in the
right panel of Fig. \ref{regions}, where the central 2\arcmin, the
arms and the shock have been excluded. The resulting power spectra for
the 1--3.5 and 3.5--7.5 keV band images are shown in the right panel
of Fig. \ref{sections} \citep[see also][]{zhuravleva14b}. We consider only scales
$\sim$100\arcsec\ and larger, i.e. below $k=10^{-2}$ arcsec$^{-1}$ 
since in this region the surface brightness is very low in the hard
band and the power drops an order of magnitude below the Poisson noise
level slightly above these scales. Around $k=10^{-2}$ arcsec$^{-1}$, both $C$ and $R$ are about 0.9, pointing to
a mixture of the three types of processes. For mean temperature $T_0
\sim 2$ keV, the values $R\sim0.9$ and $C\sim0.9$ imply relative
amplitudes of $\sim$0.4, 0.4 and 0.8 produced by adiabatic, isobaric and
isothermal processes, respectively. In terms of variance, the
majority, about 70\%, can be attributed to isothermal processes with
the remaining $\sim 30\%$ distributed equally between adiabatic and
isobaric processes.  For a lower temperature $T_0 \sim 1.6$ keV the
isothermal processes would have an even higher fraction of the total
variance, although a temperature of 1.6 keV is probably too low for
this outer region and the first estimate is more accurate. { Since this region is larger than the previous two, we can also focus on fluctuations on larger scales. At $k=3\times10^{-3}$arcsec$^{-1}$, or correspondingly spatial scales of $\sim 30$ kpc, both the coherence and the amplitude ratio drop, pointing to an increase in the relative amplitude of isobaric fluctuations. }

Once the prominent structures such as the arms and weak shocks are
excluded, the outer region appears to be dominated by isothermal
fluctuations, at least for fluctuations on scales of $\sim10$ kpc. In our analysis, isothermal structures simply correspond
to density fluctuations that produce equal amplitude fluctuations
in the soft and hard band fluxes, such as, e.g., X-ray cavities produced by
radio bubbles, or the global asymmetry of the cluster. The dominance
of isothermal fluctuations of these scales in this region is consistent with the
presence of several bubbles \citep[see, e.g.,][]{owen00,forman07}.
In addition to these bubbles, at distances of about 6\arcmin .7 from
the centre the gas is likely disturbed by large scale sloshing
\citep[e.g. Fig.~2 in][]{simionescu10,werner15}.The most prominent cold front
presumably associated with sloshing is located $\sim$20\arcmin\ north
of M87. This feature is outside the region studied here. There is also
a feature $\sim$6\arcmin\ south of M87 (green area in the lower left
quadrant of the right panel in Fig. \ref{regions}) that might be a
fainter counterpart of the more prominent northern cold front, similar
to the structures predicted by 2D numerical simulations of gas
sloshing, shown in the bottom panel in Fig.~11 in \citet{churazov03}.

\subsection{Central 2\arcmin}
\begin{figure*}
\psfig{file=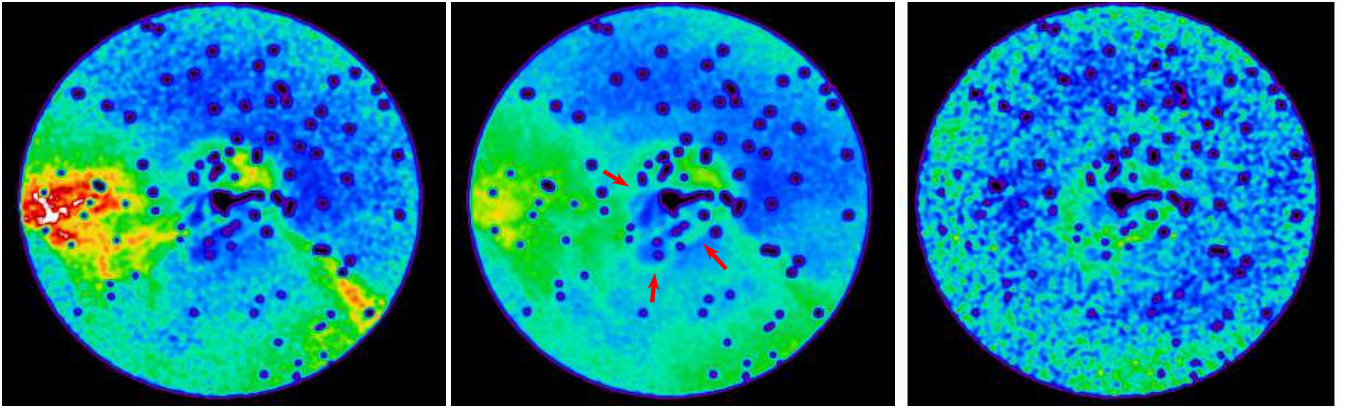,width=0.99\textwidth}
\caption{Central, 2\arcmin\ radius region, from left to right: 0.5--1,
  1--3.5 and 3.5--7.5 keV images, each divided by its best-fitting
  $\beta$-model. Note how the fluctuations corresponding to the base
  of the arms decrease with increasing energy while the ring structure
  (marked with red arrows in the central plot) becomes more dominant
  with respect to other feature with increasing energy. These trends
  are characteristic of isobaric and adiabatic fluctuations,
  respectively.}
\label{images_center}
\end{figure*}

The central 2\arcmin\ (radius) area appears disturbed by several
processes so we study it separately. Fig. \ref{images_center} shows
this region in three energy bands: 0.5--1, 1--3.5 and 3.5--7.5
keV. Each image has been divided by its best-fitting $\beta$-model to
remove the symmetric large scale structure and highlight the
fluctuations, which are shown on the same colour scale. Almost all the
bright features, especially the base of the arms toward the east and
south-west, appear stronger in the softest band, indicative of their 
isobaric nature. A few areas, marked by the red arrows, however, appear
brighter in the 1--3.5 band than in the 0.5--1 keV band, which points
towards adiabatic fluctuations. These regions include the ``bud"
structure identified by \citet{forman05,forman07} as a budding radio bubble. In the hard 3.5--7.5 keV band image (right
panel of Fig. \ref{images_center}), only this inner ring structure is
visible, suggesting an adiabatic nature of the rings and corroborating the
isobaric nature of the base of the arms.

\begin{figure}
\psfig{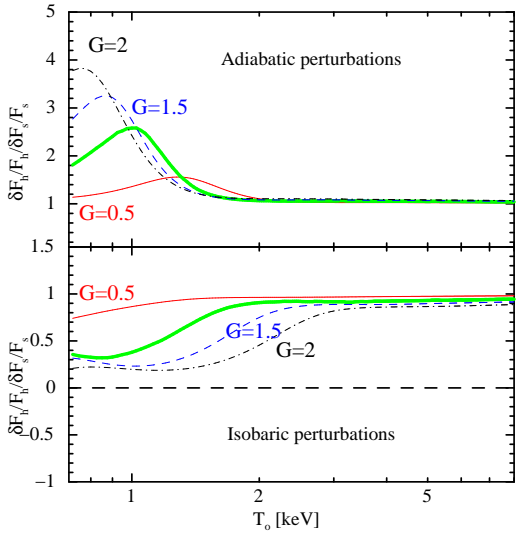}
\caption{Amplitude ratios for 0.5--1 keV and 1--3.5 keV band images
  for adiabatic (top panel) and isobaric (bottom panel) processes,
  respectively.  The fluctuations are modelled as density
  variations from an initial value $\rho_o$ by a factor $G$
  ($\rho=\rho_o*G)$. At temperatures below 3 keV, these bands have
  amplitude ratios significantly different from unity.  The adiabatic
  fluctuations appear significantly larger in the hard band, while
  isobaric fluctuations appear larger in the soft band. The thick
  green lines in each plot correspond to small amplitude fluctuations,
  while the thin lines correspond to larger (nonlinear) fluctuations
  as labeled in the figure.}
\label{dfsoft}
\end{figure}

The central region is cooler than the rest of the cluster, as can be
seen in the temperature profile in Fig. \ref{fig:tprof}, averaging
about 1.6 keV. For this low temperature, there are noticeable
differences in the response to density fluctuations in the 0.5--1 keV
and 1--3.5 keV bands, which we can exploit to study the origin of the
fluctuations further. Figure \ref{dfsoft} shows the amplitude ratios
for adiabatic and isobaric density fluctuations, similarly to
Fig. \ref{df} but this time comparing the 0.5--1 keV and 1--3.5 keV
bands. For unperturbed temperatures below 2 keV, the fractional
amplitudes deviate strongly from unity, especially for isobaric
processes.

\begin{figure}
\psfig{file=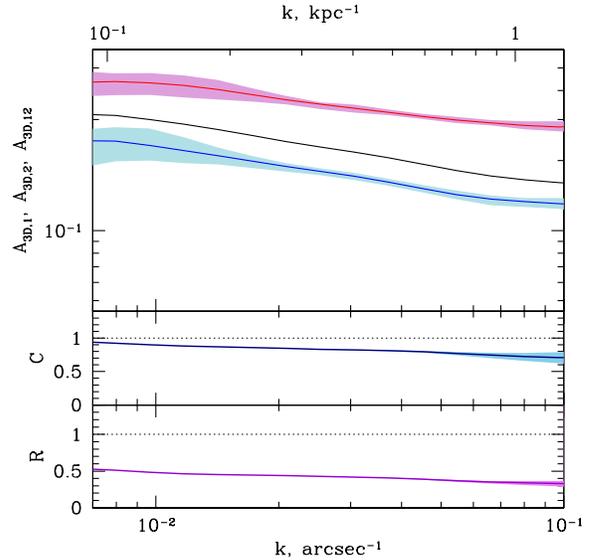,width=0.45\textwidth}
\caption{Top: Amplitude (red, blue) and cross spectrum (black) between  0.5--1 keV and 1--3.5 keV images, for the central region (left and  central images in Fig. \ref{images_center}). Middle: correlation
  coefficient, $C$. Bottom: amplitude ratio
  $R=P_{1,2}/P_1$. $R=0.5$ is consistent with mainly isobaric
  fluctuations, where the density is lower than ambient by 10\% if the
  ambient temperature is $T_0$=1.2 keV (green line in
  Fig. \ref{dfsoft}), by a factor of 1.5 from an ambient temperature of
  1.8 keV or by a factor of 2 from a ambient temperature of 2 keV
  (green, blue and black lines, respectively, in
  Fig. \ref{dfsoft}). The correlation coefficient $C$ ranging from 0.7
  to 0.9 is consistent with dominant isobaric fluctuations,
  contaminated by other processes.}
\label{center_soft}
\end{figure}

The power spectrum analysis of the central region for the very soft
0.5--1 keV and soft 1--3.5 keV energy bands is shown in
Fig. \ref{center_soft}. The correlation coefficient is $C\sim 0.9-0.7$
and $R\sim 0.5-0.4$. These value can be compared directly with the
curves in Fig. \ref{dfsoft}. For an unperturbed temperature $T_0= 2$
keV, an amplitude ratio of 0.5 requires a factor of $\sim$2 density
increase due to isobaric processes. If the unperturbed temperature is
$T_0= 1.6$, then the required density increase $\sim 1.5$ is more
modest.  The correlation coefficient drops slightly towards smaller
spatial scales (higher values of $k$) hinting at uncorrelated
fluctuations of a different type becoming stronger at smaller scales,
probably the adiabatic inner ring structure marked with red arrows in
Fig. \ref{images_center}.

\begin{figure}
\psfig{file=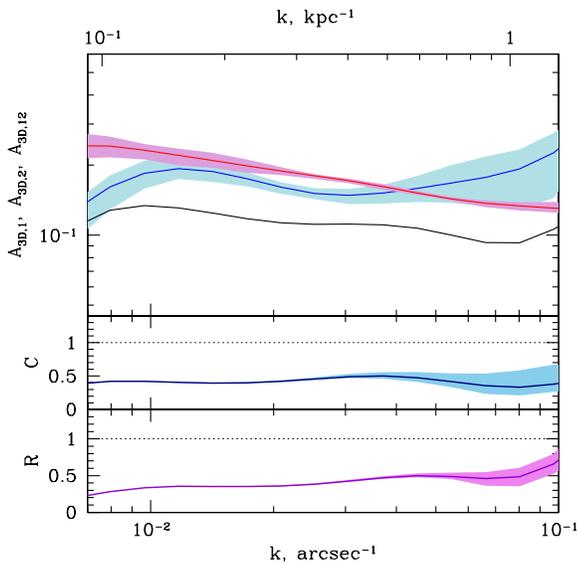,width=0.45\textwidth}
\caption{Top: Amplitude (red, blue) and cross spectrum (black) between 1--3.5 keV and 3.5--7.5 keV images, for the central  region (central and right images in Fig. \ref{images_center}). Middle:  correlation coefficient $C$. Bottom: amplitude ratio $R=P_{1,2}/P_1$. On scales of 2\arcmin or equivalently 9 kpc, $C=0.4$ and $R= 0.3$, which correspond to the top left corner of the $R-C$ maps in Fig. \ref{coh_ratio}, i.e. mostly isobaric fluctuations. On scales of 20\arcsec , about the thickness of the ring in the right panel in Fig.  \ref{images_center}, the values of $R$ and $C$ both increase to 0.5, corresponding to an increased contribution from adiabatic processes which make up about 30\% of the variance of density fluctuations at these smaller scales.}
\label{center}
\end{figure}

For this region, we have also calculated the values of $R$ and $C$ for
the images in the 1--3.5 and 3.5--7.5 keV energy bands (see
Fig.~\ref{center}).  On scales $\sim$2\arcmin\ (or equivalently
$\sim$9 kpc), similar to the radius of this central region, $C\sim
0.4$ and $R\sim 0.3$, which correspond to the top left corner of the
$R-C$ maps in Fig. \ref{coh_ratio}, i.e. mostly isobaric fluctuations,
confirming the result of the low energy bands. On scales of
$\sim$20\arcsec , about the thickness of the ring in the right panel
in Fig.  \ref{images_center}, the values of $R$ and $C$ both increase
to $\sim 0.5$, corresponding to an increased contribution from
adiabatic processes which make up about 30\% of the variance of
density fluctuations at these smaller scales.

\section{Energy content of the fluctuations}
\label{sec:discussion}

We now relate the total excess energy, associated with fluctuations,
to the observed integrated variance $M^2$. Consider, for example, the
density fluctuations produced by a collection of bubbles of
relativistic plasma, occupying a fraction $f_b\ll 1$ of the cluster
volume. We assume that the bubbles are completely devoid from thermal
plasma. Therefore, in X-ray observations they correspond to density
fluctuations $\delta\rho/\langle \rho \rangle\sim -1$. The resulting
total variance (to leading order in $f_b$) is
\begin{eqnarray}
  M^2=\left   \langle \left ( \frac{\delta \rho}{\langle \rho \rangle}\right )^2\right \rangle\sim f_b.
\end{eqnarray}
Since the free energy, associated with bubbles is their enthalpy
$E_b=\frac{\gamma}{\gamma-1}PV_b$ \citep{churazov01}, we can write
$E_b=\gamma f_b E_{\rm thermal}$, where $V_b$ is the volume occupied by
bubbles and $E_{\rm thermal}=\frac{1}{\gamma-1}PV$ is the total energy in the volume of the region under consideration, $V$. Thus
\begin{eqnarray}
\frac{E_b}{E_{\rm thermal}}\sim \gamma M^2.
\end{eqnarray}
Similar expressions can be derived for density fluctuations due to
sound waves or gravity waves, they all have the form
$\frac{E_{\rm fluctuation}}{E_{\rm thermal}}\sim \alpha M^2$, where $\alpha$
is of order unity (Zhuravleva et al. in prep). Therefore, independently of
the nature of fluctuations, we can crudely estimate the total energy
associated with fluctuations as $\approx M^2$.

To estimate the total variance of density fluctuations, $M^2$, we will
use the power or amplitude spectra of the 1--3.5 keV images, since
these correspond most closely to density fluctuations. In terms of
the power this reads
\begin{equation}
M^2=\int^{k_2}_{k_1}P_{3D}(k)4\pi k^2 dk
\end{equation}
where $P_{3D}(k)$ is the de-projected power spectrum and $k$ is the spatial wave number. In terms of the 3D amplitude of fluctuations, as plotted in the top panels in Figs. \ref{mask1} and \ref{sections} this equates to
\begin{equation}
M^2=\int^{k_2}_{k_1}\frac{A^2_{3D}(k)}{k} dk.
\label{eq:M}
\end{equation}

We integrated the 3D amplitude spectra obtained for the entire
$14\arcmin \times14\arcmin$ region in the 1--3.5 keV band, between
$k_1=0.005$ arcsec$^{-1}$ and $k_2=0.05$ arcsec$^{-1}$ obtaining a
value of $M^2\sim 0.05$. This limited range in spatial scales used for
the integration (see Eq. \ref{eq:M}) evidently results in a lower
limit to the total variance of density fluctuations. We do not expect
a large contribution to the variance from larger $k$ since the
amplitude spectrum declines with increasing $k$. On the other end,
$k<k_1$, corresponding to larger scales, could potentially provide a
significant contribution to the variance. However, the value of $k_1$
is not far from $1/R$ where $R$ is the size of the region. It is
therefore unlikely that our estimates of the total variance at scales
$<R$ are underestimated by a large factor.

For the $14\arcmin\times 14\arcmin$ region, $M^2\sim 0.05$ as discussed
above, implying that in this region the amount of non-thermal energy
is $\sim 5\%$ of the total thermal energy. From here we can estimate
that the time $t_{\rm diss}$ needed to convert this non-thermal energy
into heat in order to balance the cooling losses is $t_{\rm
  diss}=t_{\rm cooling} \times E_{\rm fluctuation}/E_{\rm thermal}\sim
0.05~t_{\rm cooling}$. For typical conditions in cool-core clusters,
this corresponds to the dynamic timescale, within a factor of a
few. This timescale is consistent with the AGN feedback model
\citep{churazov00, churazov01} when bubbles of relativistic plasma
initially capture much of the AGN mechanical power and then transfer
energy to the gas on times scales set by the rise of buoyant
bubbles. According to this model, the energy goes into turbulence and
gravity waves ---including the contribution of entrained gas--- and
dissipates into heat.

A similar analysis can be done for the amplitude spectra of selected
regions, as shown in Figs. \ref{regions} and \ref{sections}. In this
case we took two approaches. First, we integrate the amplitude spectrum
directly, in a range of scales where this spectrum is reliably
measured. Then we used this  range of $k$ to fit a powerlaw
model to the amplitude spectra and integrated the model over the
same range of $k$ as used above for the entire image
i.e. $k=0.005-0.05$arcsec$^{-1}$. The results, with both methods in
each region, are summarised in Table \ref{powers}. The $M^2$ in the
arms region is about four times higher than for the entire image,
which has the arms contribution to the power diluted by the less
perturbed part of the image. The shock and the outer regions have much
lower amplitudes, with $M^2$ about an order of magnitude lower than
the arms region.

\begin{table}
\begin{tabular}{c | c c | c c} 
Region&$k$ range&$M^2_{\rm direct}$ & $k$ range & $M^2_{\rm fit}$\\ 
\hline 
all&0.005--0.05&0.058&0.005--0.05&0.053 \\
center&0.007--0.1 &0.11 &0.005--0.05&0.098 \\
arms & 0.01--0.1& 0.18&0.005--0.05&0.22\\
shock &0.007--0.04&0.016&0.005--0.05&0.019\\
outer & 0.005--0.02&0.012&0.005--0.05&0.015\\
\hline
\end{tabular}
\caption{\rm Integrated power of density fluctuations for the regions
  noted in the left column. Direct integration of the amplitude
  spectra are quoted in the third column and analytic integration of
  the powerlaw fit to the spectra are quoted in the fifth column. The
  frequency ranges used for each integration are quoted in the second
  and fourth columns.}
\label{powers}
\end{table}

We can estimate the contribution of each section of the image to the
total power as follows. The outer region, where the shocks and arms
are excluded, is probably representative of the average conditions in
the cluster, i.e. similar bubbles and other structures should also
exist in the image regions that have been masked out, perhaps at
different position along the line of sight than the arms and
shocks. Therefore, the contribution to the total $M^2$ of these mainly
isothermal structures is likely similar to the $M^2\sim 0.015$
contribution found in the outer region. The shock region is probably
representative of the entire ring with inner radius 1\arcmin .3 and
outer radius 4\arcmin .5 from the centre, not just the section used
for our calculations, that masked out the overlying arms and other
structures. Therefore, if no other shocks are present, the shock will
contribute with about 0.02 $\times0.3=0.006$ to the total $M^2$ where
the factor of 0.3 accounts for the ratio of areas of the shock ring to
the entire $14\times14$ image. For the arms region, if these are the
only such structures in the total image then their contribution only
needs to be scaled down by the ratio of areas so that the contribution
of arms to the total $M^2$ is $\sim 0.22\times 1/9\sim0.024$. Finally,
the central region, scaled down by the area ratio contributes with
$\sim 0.1 \times 1/15= 0.006$, which in total add up approximately to
the variance of the entire region, i.e. $M^2\sim0.05$.

Note that the dissipation timescale discussed above corresponds to an
average time over the entire $14\times 14\arcmin$ region. Isolating
the central region, within 2\arcmin\ from the centre, the corresponding
$M^2 \sim 0.1$ so here $t_{\rm diss}\sim 0.1 ~t_{\rm
  cooling}$. Repeating these calculations for radii up to
4\arcmin\ also produces $t_{\rm diss}\sim 0.1~ t_{\rm cooling}$.

\section{Conclusions}
\label{sec:conclusion}

 In this paper, we demonstrate how adiabatic, isobaric and isothermal
 fluctuations in the ICM have different impacts on cluster images in
 different X-ray bands. For our chosen energy bands (1--3.5 and
 3.5--7.5 keV) adiabatic fluctuations, such as sound waves or weak
 shocks, produce larger amplitude in the hard band. On the contrary,
 isobaric fluctuations, characteristic of subsonic gas displacement,
 produce larger surface brightness fluctuations in the soft band,
 while the hard band image might show only a small or even a negative
 response to the same fluctuations. Isothermal fluctuations, such as
 bubbles and large scale asymmetries, appear equally strong in both
 bands. We have exploited this behaviour of the cluster X-ray emission
 to study the origin of the features in the ICM that produce the
 observed surface brightness fluctuations. Of course, there are a
 number of caveats, associated with this analysis. These include the
 uncertainty in removal of the global cluster profile using a simplistic
 $\beta$-model; the assumption of small-amplitude linear fluctuations,
 and the assumption that processes of different type are spatially
 uncorrelated. We nevertheless believe that the analysis captures
 important observational signatures of different processes responsible
 for the observed fluctuations and allows us to identify the most
 prominent ones.

The most prominent features of the M87 X-ray images had been
identified as cool arms and quasi-spherical shocks \citep[see
  figs. 2a, 9a, 11a in][]{forman07}. We isolated the regions that contain these structures and calculated power and
cross-power spectra, the correlation coefficient $C$ and the ratio of
amplitudes $R$. The derived values of $R$ and $C$ show an excellent
agreement with the predictions of isobaric fluctuations in the arms
region and adiabatic fluctuations in the shock region.

The rest of the cluster is not clearly dominated by a single type of
fluctuation. We divided the remaining area into a central
2\arcmin\ circle and an outer region. In the centre, the leading type
of fluctuation is isobaric, dominating the variance on scales of
100\arcsec . At smaller scales of 20\arcsec , adiabatic fluctuations
appear and account for about 30\% of the variance. The dominance
  of isobaric structures in the central region is further confirmed by
  the comparison of the soft 1--3.5 keV to the very soft 0.5--1 keV
  images (see left and centre panels in Fig. \ref{images_center}).
  The amplitude ratio $R$ between these bands is about 0.5 which is
  consistent with a factor of $\sim$2 overdensity region  in pressure
  equilibrium with ambient gas at about 1.6 keV, as shown by the
  predicted amplitude ratios for isobaric fluctuations between these
  bands in Fig. \ref{dfsoft}. 

{ For spatial scales of 5--10 kpc,} the outer region is dominated by isothermal
fluctuations, which include bubbles and deviations from the
spherically-symmetric model used to remove the large scale gradient
produced by the potential of the cluster. The measured $R$ and $C$
values correspond to about 70\% of the variance produced by isothermal
processes and the remainder is distributed about equally between
adiabatic and isobaric processes.  Fluctuations on larger scales, of about 30 kpc point to a larger contribution from isobaric fluctuations. Part of these isobaric and
apparently adiabatic fluctuations can be unrelated to the central AGN
activity, for example as a result of gas sloshing produced by
in-falling substructure.

We show that the energy content of the density fluctuations is
proportional to their integrated variance, so that by integrating the
amplitude spectra of the 1--3.5 keV images, we can estimate the ratio
of non-thermal (fluctuations) energy to the thermal energy. Using
this argument ,we find a non-thermal energy fraction
$E_{\rm fluctuation}/E_{\rm thermal}\sim 0.05$ for the entire
$14\arcmin\times 14\arcmin$ region centered on M87. From this value we
estimate the dissipation time of the fluctuations as $t_{\rm diss}\sim
0.05 t_{\rm cooling}$ in order for these fluctuations to balance
radiative cooling. The total variance depends on the region of the
image considered, in particular, in the central 2\arcmin\ radius and
the central 4\arcmin\ radius regions the integrated variance is
$M^2\sim 10\%$, indicating shorter dissipation timescales of $t_{\rm
  diss}\sim 0.1 t_{\rm cooling}$. These results broadly agree with an
AGN feedback model mediated by bubbles of relativistic plasma.

\section*{acknowledgments}
We are grateful to Dr. A. Schekochihin for many useful discussions. P.A. acknowledges support from Fondecyt grant 1140304. W.F. acknowledges support from the Smithsonian Astrophysical Observatory and NASA-Chandra archive proposal AR1-12007X.

\bibliographystyle{apj}

\begin{thebibliography}{33}
\expandafter\ifx\csname natexlab\endcsname\relax\def\natexlab#1{#1}\fi

\bibitem[{{Ar{\'e}valo} {et~al.}(2012){Ar{\'e}valo}, {Churazov}, {Zhuravleva},
  {Hern{\'a}ndez-Monteagudo}, \& {Revnivtsev}}]{rmsk}
{Ar{\'e}valo}, P., {Churazov}, E., {Zhuravleva}, I.,
  {Hern{\'a}ndez-Monteagudo}, C., \& {Revnivtsev}, M. 2012, \mnras, 426, 1793

\bibitem[{{B{\^i}rzan} {et~al.}(2004){B{\^i}rzan}, {Rafferty}, {McNamara},
  {Wise}, \& {Nulsen}}]{birzan04}
{B{\^i}rzan}, L., {Rafferty}, D.~A., {McNamara}, B.~R., {Wise}, M.~W., \&
  {Nulsen}, P.~E.~J. 2004, \apj, 607, 800

\bibitem[{{Bohringer} {et~al.}(1995){Bohringer}, {Nulsen}, {Braun}, \&
  {Fabian}}]{bohringer95}
{Bohringer}, H., {Nulsen}, P.~E.~J., {Braun}, R., \& {Fabian}, A.~C. 1995,
  \mnras, 274, L67

\bibitem[{{Churazov} {et~al.}(2001){Churazov}, {Br{\"u}ggen}, {Kaiser},
  {B{\"o}hringer}, \& {Forman}}]{churazov01}
{Churazov}, E., {Br{\"u}ggen}, M., {Kaiser}, C.~R., {B{\"o}hringer}, H., \&
  {Forman}, W. 2001, \apj, 554, 261

\bibitem[{{Churazov} {et~al.}(2000){Churazov}, {Forman}, {Jones}, \&
  {B{\"o}hringer}}]{churazov00}
{Churazov}, E., {Forman}, W., {Jones}, C., \& {B{\"o}hringer}, H. 2000, \aap,
  356, 788

\bibitem[{{Churazov} {et~al.}(2003){Churazov}, {Forman}, {Jones}, \&
  {B{\"o}hringer}}]{churazov03}
{Churazov}, E., {Forman}, W., {Jones}, C., \& {B{\"o}hringer}, H. 2003, \apj, 590, 225

\bibitem[{{Churazov} {et~al.}(2008){Churazov}, {Forman}, {Vikhlinin},
  {Tremaine}, {Gerhard}, \& {Jones}}]{churazov08}
{Churazov}, E., {Forman}, W., {Vikhlinin}, A., {Tremaine}, S., {Gerhard}, O.,
  \& {Jones}, C. 2008, \mnras, 388, 1062

\bibitem[{{Churazov} {et~al.}(2002){Churazov}, {Sunyaev}, {Forman}, \&
  {B{\"o}hringer}}]{2002MNRAS.332..729C}
{Churazov}, E., {Sunyaev}, R., {Forman}, W., \& {B{\"o}hringer}, H. 2002,
  \mnras, 332, 729

\bibitem[{{Churazov} {et~al.}(2012){Churazov}, {Vikhlinin}, {Zhuravleva},
  {Schekochihin}, {Parrish}, {Sunyaev}, {Forman}, {B{\"o}hringer}, \&
  {Randall}}]{churazov12}
{Churazov}, E., {Vikhlinin}, A., {Zhuravleva}, I., {Schekochihin}, A.,
  {Parrish}, I., {Sunyaev}, R., {Forman}, W., {B{\"o}hringer}, H., \&
  {Randall}, S. 2012, \mnras, 421, 1123

\bibitem[{{David} {et~al.}(2001){David}, {Nulsen}, {McNamara}, {Forman},
  {Jones}, {Ponman}, {Robertson}, \& {Wise}}]{david01}
{David}, L.~P., {Nulsen}, P.~E.~J., {McNamara}, B.~R., {Forman}, W., {Jones},
  C., {Ponman}, T., {Robertson}, B., \& {Wise}, M. 2001, \apj, 557, 546

\bibitem[{{Dunn} \& {Fabian}(2006)}]{dunn06}
{Dunn}, R.~J.~H. \& {Fabian}, A.~C. 2006, \mnras, 373, 959

\bibitem[{{Fabian} {et~al.}(2003){Fabian}, {Sanders}, {Allen}, {Crawford},
  {Iwasawa}, {Johnstone}, {Schmidt}, \& {Taylor}}]{2003MNRAS.344L..43F}
{Fabian}, A.~C., {Sanders}, J.~S., {Allen}, S.~W., {Crawford}, C.~S.,
  {Iwasawa}, K., {Johnstone}, R.~M., {Schmidt}, R.~W., \& {Taylor}, G.~B. 2003,
  \mnras, 344, L43

\bibitem[{{Fabian} {et~al.}(2000){Fabian}, {Sanders}, {Ettori}, {Taylor},
  {Allen}, {Crawford}, {Iwasawa}, {Johnstone}, \& {Ogle}}]{fabian00}
{Fabian}, A.~C., {Sanders}, J.~S., {Ettori}, S., {Taylor}, G.~B., {Allen},
  S.~W., {Crawford}, C.~S., {Iwasawa}, K., {Johnstone}, R.~M., \& {Ogle}, P.~M.
  2000, \mnras, 318, L65

\bibitem[{{Forman} {et~al.}(2007){Forman}, {Jones}, {Churazov}, {Markevitch},
  {Nulsen}, {Vikhlinin}, {Begelman}, {B{\"o}hringer}, {Eilek}, {Heinz},
  {Kraft}, {Owen}, \& {Pahre}}]{forman07}
{Forman}, W., {Jones}, C., {Churazov}, E., {Markevitch}, M., {Nulsen}, P.,
  {Vikhlinin}, A., {Begelman}, M., {B{\"o}hringer}, H., {Eilek}, J., {Heinz},
  S., {Kraft}, R., {Owen}, F., \& {Pahre}, M. 2007, \apj, 665, 1057

\bibitem[{{Forman} {et~al.}(2005){Forman}, {Nulsen}, {Heinz}, {Owen}, {Eilek},
  {Vikhlinin}, {Markevitch}, {Kraft}, {Churazov}, \& {Jones}}]{forman05}
{Forman}, W., {Nulsen}, P., {Heinz}, S., {Owen}, F., {Eilek}, J., {Vikhlinin},
  A., {Markevitch}, M., {Kraft}, R., {Churazov}, E., \& {Jones}, C. 2005, \apj,
  635, 894

\bibitem[{{Harris} {et~al.}(2003){Harris}, {Biretta}, {Junor}, {Perlman},
  {Sparks}, \& {Wilson}}]{harris03}
{Harris}, D.~E., {Biretta}, J.~A., {Junor}, W., {Perlman}, E.~S., {Sparks},
  W.~B., \& {Wilson}, A.~S. 2003, \apjl, 586, L41

\bibitem[{{Jord{\'a}n} {et~al.}(2004){Jord{\'a}n}, {C{\^o}t{\'e}}, {Ferrarese},
  {Blakeslee}, {Mei}, {Merritt}, {Milosavljevi{\'c}}, {Peng}, {Tonry}, \&
  {West}}]{jordan04}
{Jord{\'a}n}, A., {C{\^o}t{\'e}}, P., {Ferrarese}, L., {Blakeslee}, J.~P.,
  {Mei}, S., {Merritt}, D., {Milosavljevi{\'c}}, M., {Peng}, E.~W., {Tonry},
  J.~L., \& {West}, M.~J. 2004, \apj, 613, 279

\bibitem[{{Marshall} {et~al.}(2002){Marshall}, {Miller}, {Davis}, {Perlman},
  {Wise}, {Canizares}, \& {Harris}}]{marshall02}
{Marshall}, H.~L., {Miller}, B.~P., {Davis}, D.~S., {Perlman}, E.~S., {Wise},
  M., {Canizares}, C.~R., \& {Harris}, D.~E. 2002, \apj, 564, 683

\bibitem[{{McNamara} {et~al.}(2000){McNamara}, {Wise}, {Nulsen}, {David},
  {Sarazin}, {Bautz}, {Markevitch}, {Vikhlinin}, {Forman}, {Jones}, \&
  {Harris}}]{2000ApJ...534L.135M}
{McNamara}, B.~R., {Wise}, M., {Nulsen}, P.~E.~J., {David}, L.~P., {Sarazin},
  C.~L., {Bautz}, M., {Markevitch}, M., {Vikhlinin}, A., {Forman}, W.~R.,
  {Jones}, C., \& {Harris}, D.~E. 2000, \apjl, 534, L135

\bibitem[{{Million} {et~al.}(2010){Million}, {Werner}, {Simionescu}, {Allen},
  {Nulsen}, {Fabian}, {B{\"o}hringer}, \& {Sanders}}]{million10}
{Million}, E.~T., {Werner}, N., {Simionescu}, A., {Allen}, S.~W., {Nulsen},
  P.~E.~J., {Fabian}, A.~C., {B{\"o}hringer}, H., \& {Sanders}, J.~S. 2010,
  \mnras, 407, 2046

\bibitem[{{Molendi}(2002)}]{molendi02}
{Molendi}, S. 2002, \apj, 580, 815

\bibitem[{{Nulsen} {et~al.}(2002){Nulsen}, {David}, {McNamara}, {Jones},
  {Forman}, \& {Wise}}]{nulsen02}
{Nulsen}, P.~E.~J., {David}, L.~P., {McNamara}, B.~R., {Jones}, C., {Forman},
  W.~R., \& {Wise}, M. 2002, \apj, 568, 163

\bibitem[{{Owen} {et~al.}(2000){Owen}, {Eilek}, \& {Kassim}}]{owen00}
{Owen}, F.~N., {Eilek}, J.~A., \& {Kassim}, N.~E. 2000, \apj, 543, 611

\bibitem[{{Randall} {et~al.}(2011){Randall}, {Forman}, {Giacintucci}, {Nulsen},
  {Sun}, {Jones}, {Churazov}, {David}, {Kraft}, {Donahue}, {Blanton},
  {Simionescu}, \& {Werner}}]{2011ApJ...726...86R}
{Randall}, S.~W., {Forman}, W.~R., {Giacintucci}, S., {Nulsen}, P.~E.~J.,
  {Sun}, M., {Jones}, C., {Churazov}, E., {David}, L.~P., {Kraft}, R.,
  {Donahue}, M., {Blanton}, E.~L., {Simionescu}, A., \& {Werner}, N. 2011,
  \apj, 726, 86

\bibitem[{{Reynolds} {et~al.}(2005){Reynolds}, {McKernan}, {Fabian}, {Stone},
  \& {Vernaleo}}]{2005MNRAS.357..242R}
{Reynolds}, C.~S., {McKernan}, B., {Fabian}, A.~C., {Stone}, J.~M., \&
  {Vernaleo}, J.~C. 2005, \mnras, 357, 242

\bibitem[{{Simionescu} {et~al.}(2010){Simionescu}, {Werner}, {Forman},
  {Miller}, {Takei}, {B{\"o}hringer}, {Churazov}, \& {Nulsen}}]{simionescu10}
{Simionescu}, A., {Werner}, N., {Forman}, W.~R., {Miller}, E.~D., {Takei}, Y.,
  {B{\"o}hringer}, H., {Churazov}, E., \& {Nulsen}, P.~E.~J. 2010, \mnras, 405,
  91

\bibitem[{{Vikhlinin} {et~al.}(2005){Vikhlinin}, {Markevitch}, {Murray},
  {Jones}, {Forman}, \& {Van Speybroeck}}]{vikhlinin05}
{Vikhlinin}, A., {Markevitch}, M., {Murray}, S.~S., {Jones}, C., {Forman}, W.,
  \& {Van Speybroeck}, L. 2005, \apj, 628, 655

\bibitem[{{Werner} {et~al.}(2015){Werner}, {ZuHone}, {Zhuravleva}, {Ichinohe},
  {Simionescu}, {Allen}, {Markevitch}, {Fabian}, {Keshet}, {Roediger},
  {Ruszkowski}, \& {Sanders}}]{werner15}
{Werner}, N., {ZuHone}, J.~A., {Zhuravleva}, I., {Ichinohe}, Y., {Simionescu},
  A., {Allen}, S.~W., {Markevitch}, M., {Fabian}, A.~C., {Keshet}, U.,
  {Roediger}, E., {Ruszkowski}, M., \& {Sanders}, J.~S. 2015, ArXiv e-prints

\bibitem[{{Young} {et~al.}(2002){Young}, {Wilson}, \& {Mundell}}]{young02}
{Young}, A.~J., {Wilson}, A.~S., \& {Mundell}, C.~G. 2002, \apj, 579, 560

\bibitem[{{Zhuravleva} {et~al.}(2015){Zhuravleva}, {Churazov}, {Ar'evalo},
  {Schekochihin}, {Allen}, {Fabian}, {Forman}, {Sanders}, {Simionescu},
  {Sunyaev}, {Vikhlinin}, \& {Werner}}]{zhuravleva15}
{Zhuravleva}, I., {Churazov}, E., {Ar'evalo}, P., {Schekochihin}, A.~A.,
  {Allen}, S.~W., {Fabian}, A.~C., {Forman}, W.~R., {Sanders}, J.~S.,
  {Simionescu}, A., {Sunyaev}, R., {Vikhlinin}, A., \& {Werner}, N. 2015, ArXiv
  e-prints

\bibitem[{{Zhuravleva} {et~al.}(2014){Zhuravleva}, {Churazov}, {Schekochihin},
  {Allen}, {Ar{\'e}valo}, {Fabian}, {Forman}, {Sanders}, {Simionescu},
  {Sunyaev}, {Vikhlinin}, \& {Werner}}]{zhuravleva14b}
{Zhuravleva}, I., {Churazov}, E., {Schekochihin}, A.~A., {Allen}, S.~W.,
  {Ar{\'e}valo}, P., {Fabian}, A.~C., {Forman}, W.~R., {Sanders}, J.~S.,
  {Simionescu}, A., {Sunyaev}, R., {Vikhlinin}, A., \& {Werner}, N. 2014, \nat,
  515, 85

\end{thebibliography}

\end{document}